\documentclass[12pt,a4paper]{article}
\usepackage{jheppub}
\usepackage[usenames,dvipsnames]{xcolor}
\usepackage{amssymb} 
\usepackage{amsmath}
\usepackage{amsmath}
\allowdisplaybreaks[4]
\usepackage{mathtools}
\usepackage{amsfonts}    
\usepackage{dsfont}
\usepackage{breqn}
\usepackage{pdfpages}
\usepackage{verbatim}
\usepackage{stmaryrd}
\usepackage{graphicx}
\usepackage{cancel}
\usepackage{tensor}
\usepackage{mathrsfs}
\usepackage{textgreek} 
\usepackage[mathscr]{euscript}
\usepackage[smalltableaux]{ytableau}
\ytableausetup{centertableaux}
\usepackage{tikz}
\usetikzlibrary{decorations.pathreplacing, decorations.markings,calc,shapes.misc,decorations.pathmorphing,patterns.meta, math}
\usepackage{slashed}
\usepackage{datetime}
\usepackage{hyperref}
\hypersetup{
    pdfencoding=unicode,
	colorlinks=true,
	urlcolor=Maroon,
	linkcolor=RoyalBlue,
	citecolor=Maroon,
	pdftitle={The Generating Function of Sunset},
	pdfauthor={Xinhe Chen, Bo Feng, Liang Zhang},
	pdfdisplaydoctitle=true,
	pdfstartview=FitH,
	linktocpage=true
}
 
\usetikzlibrary{shapes}

\newcommand{\ba}{\begin{align}}
\newcommand{\ea}{\end{align}}
\newcommand{\be}{\begin{equation}}
\newcommand{\ee}{\end{equation}}
\newcommand{\bea}{\begin{eqnarray}}
	\newcommand{\eea}{\end{eqnarray}}

\def\bd{\begin{tikzpicture}}
\def\ed{\end{tikzpicture}}

\allowdisplaybreaks

\def\XXint#1#2#3{{\setbox0=\hbox{$#1{#2#3}{\int}$}
     \vcenter{\hbox{$#2#3$}}\kern-.5\wd0}}

\definecolor{light-gray}{gray}{0.75}

\title{Tensor Reduction of Sunset by Generating Function}

 \author[a]{Xinhe Chen,}
 \author[b,c,d,e]{Bo Feng,}
 \author[d]{Liang Zhang}
\emailAdd{chenxinhe24@gscaep.ac.cn}
\emailAdd{fengbo@scnu.edu.cn}
\emailAdd{liangzhang@csrc.ac.cn}
 \affiliation[a]{Graduate School of China Academy of Engineering Physics, No. 10 Xibeiwang East Road, Haidian District, Beijing, 100193, China}
  \affiliation[b]{State Key Laboratory of Nuclear Physics and Technology, Institute of Quantum Matter, South China Normal University, Guangzhou 510006, China}
 \affiliation[c]{Guangdong Basic Research Center of Excellence for Structure and Fundamental Interactions of Matter, Guangdong Provincial Key Laboratory of Nuclear Science, Guangzhou 510006, China}
 \affiliation[d]{Beijing Computational Science Research Center, Beijing 100084, China}
 \affiliation[e]{Peng Huanwu Center for Fundamental Theory, Hefei, Anhui, 230026, China}

\abstract{Recently, the generating function has been proposed as an alternative reduction method. This method has been tested at the one-loop level, including the tensor reduction and propagators with higher powers. In this work, we initiate the study of the method for higher loops by focusing on the sunset diagram, which is the simplest nontrivial two-loop integral. By employing Passarino–Veltman (PV) reduction equations together with syzygy equations, we construct a complete system of differential equations. Through series expansion, we derive a complete set of recurrence relations, which can efficiently reduce any high-rank tensor structure.
}

\begin{document}

\setcounter{tocdepth}{3}
\maketitle
\setcounter{page}{2}

\section{\label{sec:intro}Introduction}
The scattering amplitude, as a core physical quantity in quantum field theory, directly encodes the quantum transition probability from an initial state to a final state. Therefore, the calculation and understanding of scattering amplitudes have always been a central task in theoretical high-energy physics research. Over the past two decades, the computation of one-loop Feynman integrals has been largely resolved. However, as experimental precision continues to improve, attention has increasingly shifted to the calculation of two and higher loop Feynman integrals, leading to the development of a series of efficient computational methods (see the books \cite{Smirnov:2012gma,Weinzierl:2022eaz}).

Currently, the mainstream methods for computing Feynman integrals are divided into two steps. The first step involves decomposing a general loop integral into a linear combination of a set of basis integrals (called master integrals), obtaining their reduction coefficients. The second step is to compute these master integrals. In recent years, the academic community has developed many methods for computing reduction coefficients, with the integration-by-parts (IBP) identities \cite{Tkachov:1981wb,Chetyrkin:1981qh} currently being the preferred approach\footnote{There are also other effective methods. For example, for one-loop integrals, there are Passarino–Veltman (PV) reduction \cite{Passarino:1978jh}, Ossola-Papadopoulos-Pittau (OPP) method \cite{Ossola:2006us,Ossola:2007ax}, unitarity cuts \cite{Bern:1994zx,Britto:2004nc,Anastasiou:2006jv,Britto:2006fc}. For high loops, the auxiliary mass flow method \cite{Liu:2017jxz,Liu:2022tji} and computational algebraic geometry methods \cite{Badger:2012dp,Zhang:2012ce,Larsen:2015ped} are also useful.}. 
This method involves taking total derivatives of various integrands to derive relationships among different integrals. Later, with the Laporta algorithm \cite{Laporta:2000dsw} and computational software such as  Air \cite{Anastasiou:2004vj}, LiteRed \cite{Lee:2013mka}, FIRE6 \cite{Smirnov:2019qkx,Smirnov:2023yhb}, Reduze \cite{vonManteuffel:2012np}, Kira \cite{Maierhofer:2017gsa,Klappert:2020nbg}, Forcer \cite{Ruijl:2017cxj}, FiniteFlow \cite{Peraro:2019svx}, NeatIBP \cite{Wu:2023upw}, Blade \cite{Guan:2024byi} and AmpRed \cite{Chen:2024xwt},  it became possible to obtain reduction coefficients for many Feynman integrals. However, with the increasing complexity of cutting-edge problems, this method introduces many unnecessary complexities. To address this, several advanced approaches have been developed to minimize the computational demands of traditional IBP methods, including finite field methods \cite{vonManteuffel:2014ixa,Peraro:2016wsq,Peraro:2019svx,Belitsky:2023qho}, block-triangular form reduction \cite{Liu:2018dmc,Guan:2019bcx,Guan:2024byi}, and AI-improved method \cite{vonHippel:2025okr,Song:2025pwy}.

For any of the above methods, what we ultimately obtain are recurrence relations from higher orders to lower orders. Recently, it has been discovered that the method of generating functions \cite{Ablinger:2014yaa,Kosower:2018obg,Feng:2022hyg,Guan:2023avw,Hu:2023mgc,Li:2024rvo,Hu:2024kch,Hu:2025spa,Hu:2025rrt} can more systematically express these  structures. With the introduction of auxiliary vectors for the tensor reduction or mass-like parameters for higher order propagators (see
\cite{Hu:2021nia,Feng:2022iuc,Feng:2022uqp,Feng:2022rwj,Li:2022cbx,Zhang:2023jzv,Guan:2023avw}), some nice structures of reduction can be established. However, currently, most articles focus on generating functions for one-loop integrals. For higher-loop integrals, since the computational complexity increases, the efficiency of the generating function method is  not very clear.

In this paper, we want to test the usefulness of generating functions by studying the first nontrivial two-loop integrals, i.e., the sunset diagrams. To establish differential equations for  generating functions,  we first employ the PV reduction method with  auxiliary vectors $R_i$. However,  for integrals beyond one loop, the PV reduction method only utilizes the tensor information of the integral and cannot yield a complete system of equations. To address this limitation, we establish syzygy equations \cite{Gluza:2010ws,Schabinger:2011dz} to find other independent and simpler differential equations, thereby obtaining a complete system.  

The paper is organized as follows. Section \ref{sec:IBP} and introduces the syzygy and PV equations. Section \ref{sec:solve} and \ref{sec:massless} present the solving procedure for general masses and massless case respectively and show that all reduction coefficients can be expressed in a finite basis. Section \ref{sec:conc} concludes with a summary and outlook. Appendix \ref{sec:Tadpole result} contains the boundary-term calculations, while Appendix \ref{sec:example} presents an example for a rank-four tensor reduction to illustrate the application of our result.

\section{\label{sec:IBP}The IBP relations}

In this paper, we will focus on the tensor reduction of general sunset diagrams. For this purpose, we define the  generating function as follows:
\begin{align}
	I_{gen}= \int \frac{d^D l_1}{i \pi^{D/2}}\frac{d^D l_2}{i \pi^{D/2}}  \frac{e ^ {l_1 \cdot R_1} e ^ {l_2 \cdot R_2}}{D_1 D_2 D_3} \,, \label{2.1}
\end{align}
where $l_i$ are loop momentums and $R_i$ are  auxiliary vectors. Here, the inverse  propagators $D_i$ are:
\begin{align}
	 D_1 = l^2_1 - M^2_1  ,~~~~~ D_2 = l^2_2 - M^2_2 ,~~~~~ D_3 = (l_1 + l_2 + K)^2 - M^2_3 \,,
\end{align}
with external momentum $K$. For general integrals, there are seven master integrals. We can choose the following seven master integrals:
\begin{align}
	&I_1= \int \frac{d^D l_1}{i \pi^{D/2}}\frac{d^D l_2}{i \pi^{D/2}}  \frac{1}{D_1 D_2 D_3} ,~~~~~~~~~~	I_2=\frac{1}{K^2} \int \frac{d^D l_1}{i \pi^{D/2}}\frac{d^D l_2}{i \pi^{D/2}}  \frac{l_1 \cdot K}{D_1 D_2 D_3} \,,\notag\\&	I_3= \frac{1}{K^2}\int \frac{d^D l_1}{i \pi^{D/2}}\frac{d^D l_2}{i \pi^{D/2}}  \frac{l_2 \cdot K}{D_1 D_2 D_3} \,,~~~	I_4= \frac{1}{(K^2)^2}\int \frac{d^D l_1}{i \pi^{D/2}}\frac{d^D l_2}{i \pi^{D/2}} \frac{ ({l_1 \cdot K})({l_2 \cdot K})}{D_1 D_2 D_3} \,,\notag\\&	I_5= \frac{1}{K^2}\int \frac{d^D l_1}{i \pi^{D/2}}\frac{d^D l_2}{i \pi^{D/2}}  \frac{1}{D_1 D_2 },~~~~ 	I_6= \frac{1}{K^2}\int \frac{d^D l_1}{i \pi^{D/2}} \frac{d^D l_2}{i \pi^{D/2}}  \frac{1}{D_1  D_3} , \,,\notag\\&	I_7= \frac{1}{K^2}\int \frac{d^D l_1}{i \pi^{D/2}}\frac{d^D l_2}{i \pi^{D/2}}  \frac{1}{ D_2 D_3} \,. \label{2.3}
\end{align}
 The choice of factor $K^2$ is for the match of mass dimension, so the basis has the same mass dimension. Then  the generating function can be expressed as
\begin{align}
I_{gen} =\vec{I} \cdot \vec{\alpha} \,,\label{2.4}
\end{align}
where master integrals have been written into the row vector $\vec{I} = (I_1 , I_2, I_3, I_4, I_5, I_6, I_7)$   and reduction coefficients have been written into the column vector $\vec{\alpha}$. Our goal is to establish the differential equation for  $\vec{\alpha}$ and solve it. 

To solve $\vec{\alpha}$, we need to establish enough differential equations. A very natural starting point is the fundamental IBP relations. However, in these relations, terms with doubled propagators will appear, thus they would introduce unnecessary complexity for our problem. To circumvent this complication, we implement algebraic geometry techniques to formulate IBP equations that are explicitly free of doubled propagator terms.

\subsection{The natural equations}

With the introduction of auxiliary vectors, we can immediately write down three natural equations according to the idea of PV-reduction \cite{Feng:2022uqp}.  The first one is:
\begin{align}
&\frac{\partial{}}{\partial{R_1}} \cdot \frac{\partial{}}{\partial{R_1}} I_{gen} =\int \frac{d^D l_1}{i \pi^{D/2}}\frac{d^D l_2}{i \pi^{D/2}} \frac{ l_1^2 e ^ {l_1 \cdot R_1} e ^ {l_2 \cdot R_2}}{D_1 D_2 D_3}\notag \\ & = M^2_1 \int \frac{d^D l_1}{i \pi^{D/2}}\frac{d^D l_2}{i \pi^{D/2}} \frac{e ^ {l_1 \cdot R_1} e ^ {l_2 \cdot R_2}}{D_1 D_2 D_3} + \int \frac{d^D l_1}{i \pi^{D/2}}\frac{d^D l_2}{i \pi^{D/2}} \frac{e ^ {l_1 \cdot R_1} e ^ {l_2 \cdot R_2}}{ D_2 D_3} \,. \label{2.5}
\end{align}
To write down the differential equations, we introduce three known functions, which correspond to the reduction of three subsectors to the master integrals given in \eqref{2.3}:
\begin{align}
\int \frac{d^D l_1}{i \pi^{D/2}}\frac{d^D l_2}{i \pi^{D/2}}\frac{e ^ {l_1 \cdot R_1} e ^ {l_2 \cdot R_2}}{D_1 D_2} & = K^2 \vec{I} \cdot \vec{\beta}_{12} \,,\nonumber  \\
\int \frac{d^D l_1}{i \pi^{D/2}}\frac{d^D l_2}{i \pi^{D/2}}\frac{e ^ {l_1 \cdot R_1} e ^ {l_2 \cdot R_2}}{D_1 D_3} & = K^2 \vec{I} \cdot \vec{\beta}_{13}\,,\nonumber \\
 \int \frac{d^D l_1}{i \pi^{D/2}}\frac{d^D l_2}{i \pi^{D/2}}\frac{e ^ {l_1 \cdot R_1} e ^ {l_2 \cdot R_2}}{ D_2 D_3} & = K^2 \vec{I} \cdot \vec{\beta}_{23}\,.\label{2.8}
\end{align}
Then from \eqref{2.4}, \eqref{2.5} and \eqref{2.8}, we can obtain the differential equation for the coefficient $\vec{\alpha}$:
\begin{align}
\frac{\partial{}}{\partial{R_1}} \cdot \frac{\partial{}}{\partial{R_1}}\vec{\alpha} = M^2_1 \vec{\alpha} + K^2\vec{\beta}_{23}\,. \label{2.9}
\end{align}	
Similarly, we obtain the second differential equation:
\begin{align}
	\frac{\partial{}}{\partial{R_2}} \cdot \frac{\partial{}}{\partial{R_2}}\vec{\alpha} = M^2_2 \vec{\alpha} + K^2\vec{\beta}_{13}\,.  \label{2.10}
\end{align}	
The third differential equation is:
 \begin{align}
\frac{\partial{}}{\partial{R_1}} \cdot \frac{\partial{}}{\partial{R_2}}\vec{\alpha}& =-(K \cdot l_1 + K \cdot l_2 + f )\vec{\alpha} + \frac{K^2}{2}(\vec{\beta}_{12}-\vec{\beta}_{23}-\vec{\beta}_{13})\notag \\ 
&=-(K \cdot \frac{\partial}{\partial R_1} + K \cdot \frac{\partial}{\partial R_2} + f )\vec{\alpha} + \frac{K^2}{2}(\vec{\beta}_{12}-\vec{\beta}_{23}-\vec{\beta}_{13})\,,\label{2.11}
\end{align}	
where the first equality employs  
 $ l_1 \cdot l_2 = - l_1 \cdot K - l_2 \cdot K + \frac{1}{2}( D_3 - D_1 - D_2) - f$ with $ f = \frac{1}{2} (K^2 + M^2_1 + M^2_2 -  M^2_3) $.

It turns out that the above three equations are not sufficient to completely determine $\vec{\alpha}$, thus it is necessary to derive additional equations to establish a complete system for solving $\vec{\alpha}$.

\subsection{The syzygy equations}

In this subsection, we will derive two other differential equations to give a complete system to determine $\vec{\alpha}$. We will closely follow the syzygy method in  \cite{Gluza:2010ws}\footnote{The method in \cite{Gluza:2010ws} is in the momentum representation (see also \cite{Zhang:2014xwa} ). The syzygy method in the Baikov representation can be found in \cite{Larsen:2015ped,Zhang:2016kfo}. }.

For general IBP relation ${\partial\over \partial \ell_i^\mu}v_i^\mu$, i.e.,
\begin{align} 
	0 = &\int \frac{d^D l_1}{i \pi^{D/2}}\frac{d^D l_2}{i \pi^{D/2}} \frac{\partial}{\partial l_1^\mu}({v_1^\mu \frac{e ^ {l_1 \cdot R_1} e ^ {l_2 \cdot R_2}}{D_1 D_2 D_3}}) + \int \frac{d^D l_1}{i \pi^{D/2}}\frac{d^D l_2}{i \pi^{D/2}} \frac{\partial}{\partial l_2^\mu}({v_2^\mu \frac{e ^ {l_1 \cdot R_1} e ^ {l_2 \cdot R_2}}{D_1 D_2 D_3}})\,.
\end{align}
We construct two vectors
\begin{align}
v_i^\mu = c^i_1 l_1^\mu + c^i_2 l_2^\mu + c^i_3 K^\mu + c^i_4 R_1^\mu + c^i_5 R_2^\mu\,,~~~~i=1,2,
\end{align}
where each of the coefficients $c^i_j$ is a function of  Lorentz scalars.
If the choices of $v_i^\mu$  satisfy the following conditions
\begin{align}
	  &\int \frac{d^D l_1}{i \pi^{D/2}}\frac{d^D l_2}{i \pi^{D/2}} v_1 \cdot \frac{\partial}{\partial l_1}(\frac{1}{D_1}) \frac{e ^ {l_1 \cdot R_1} e ^ {l_2 \cdot R_2}}{ D_2 D_3} = - \int \frac{d^D l_1}{i \pi^{D/2}}\frac{d^D l_2}{i \pi^{D/2}} 2u_1 \frac{e ^ {l_1 \cdot R_1} e ^ {l_2 \cdot  R_2}}{D_1 D_2 D_3}  \,,\notag \\
	    &\int \frac{d^D l_1}{i \pi^{D/2}}\frac{d^D l_2}{i \pi^{D/2}} v_2 \cdot \frac{\partial}{\partial l_2}(\frac{1}{D_2}) \frac{e ^ {l_1 \cdot R_1} e ^ {l_2 \cdot R_2}}{ D_1 D_3} = - \int \frac{d^D l_1}{i \pi^{D/2}}\frac{d^D l_2}{i \pi^{D/2}} 2u_2 \frac{e ^ {l_1 \cdot R_1} e ^ {l_2 \cdot  R_2}}{D_1 D_2 D_3} \,, \notag \\ 
	    &\int \frac{d^D l_1}{i \pi^{D/2}}\frac{d^D l_2}{i \pi^{D/2}} v_1 \cdot \frac{\partial}{\partial l_1}(\frac{1}{D_3}) \frac{e ^ {l_1 \cdot R_1} e ^ {l_2 \cdot R_2}}{ D_1 D_2}  +  \int \frac{d^D l_1}{i \pi^{D/2}}\frac{d^D l_2}{i \pi^{D/2}} v_2 \cdot \frac{\partial}{\partial l_2}(\frac{1}{D_3}) \frac{e ^ {l_1 \cdot R_1} e ^ {l_2 \cdot R_2}}{ D_1 D_2} \notag \\ &=-\int \frac{d^D l_1}{i \pi^{D/2}}\frac{d^D l_2}{i \pi^{D/2}} 2u_3 \frac{e ^ {l_1 \cdot R_1} e ^ {l_2 \cdot  R_2}}{D_1 D_2 D_3} \,,\label{2.14}
		\end{align}
the obtained IBP equations will be free of double propagators. Here, $u_i$ is also a function of  Lorentz scalars or simply a constant. Equation \eqref{2.14} can be recast into:
\begin{align}
	c\cdot E=0 \,,
\end{align}
where
	\begin{align}
		c &= (c^1_1,c^1_2,c^1_3,c^1_4,c^1_5,c^2_1,c^2_2,c^2_3,c^2_4,c^2_5,u_1,u_2, u_3)\,,\\
		E &= \begin{pmatrix} l^2_1 &~~~~~ 0 & ~~~~~~l_1 \cdot (l_1 + l_2 + K) \\ 
			l_1 \cdot l_2 &~~~~~  0 &~~~~~l_2 \cdot (l_1 + l_2 + K)\\ 
			K \cdot l_1 &~~~~~0 &~~~~~ K\cdot  (l_1 + l_2 + K)\\ 
			 R_1 \cdot l_1 & ~~~~~0 &~~~~~R_1  \cdot (l_1 + l_2 + K)\\ 
			 R_2 \cdot l_1  &~~~~~ 0 &~~~~~ R_2  \cdot (l_1 + l_2 + K)\\ 
			0 &~~~~~ l_1 \cdot l_2 &~~~~~l_1 \cdot (l_1 + l_2 + K)\\ 
			0 & ~~~~~l^2_2 &~~~~~l_2 \cdot (l_1 + l_2 + K)\\ 
			0 & ~~~~~K \cdot l_2 &~~~~~ K\cdot  (l_1 + l_2 + K)\\ 
			0 & ~~~~~R_1 \cdot l_2 &~~~~~R_1  \cdot (l_1 + l_2 + K)\\ 
			0 & ~~~~~R_2 \cdot l_2 &~~~~~R_2  \cdot (l_1 + l_2 + K)\\ 
			-D_1 & ~~~~~0 &~~~~~0\\ 
			0 & ~~~~~-D_2 &~~~~~0\\ 
			0 &~~~~~ 0 &~~~~~-D_3\end{pmatrix}\,.\label{2.15}
	\end{align}
Each solution to this matrix equation is called a syzygy. Using Singular \cite{DGPS}, we can access the Gröbner basis of this set of solutions, which consists of a total of 159  elements. Most of the IBP equations derived from these results are not linearly independent. To select simple IBP equations,  we want  the $R_i$-component of the vector to be zero. There are a total of thirty such vectors that can derive thirty IBP equations, and we choose two linearly independent simple equations from them.

The first pair of vectors is 
\begin{align}
v_1^\mu &= -l_1 \cdot l_2 l_1^\mu + (l_1^2 +l_1 \cdot K)l_2^\mu - l_1 \cdot l_2 K^\mu \,,\notag \\
v_2^\mu &= -(l_2^2 +l_2 \cdot K) l_1^\mu + l_1 \cdot l_2  l_2^\mu + l_1 \cdot l_2 K^\mu \,,
\end{align}
and the corresponding differential equation is
\begin{align}
0  = &\{ R_1 \cdot \frac{\partial }{\partial R_1} ( K\cdot \frac{\partial }{\partial R_1} )+  K \cdot \frac{\partial }{\partial R_2 } ( R_1 \cdot \frac{\partial }{\partial R_1}) + f  R_1 \cdot \frac{\partial }{\partial R_1} +  K \cdot R_1 ( K \cdot \frac{\partial }{\partial R_1})   \notag\\ 
&+  K \cdot R_1 ( K \cdot \frac{\partial }{\partial R_2}) + f  K \cdot R_1  + M^2_1  R_1 \cdot \frac{\partial }{\partial R_2} +  R_1 \cdot \frac{\partial }{\partial R_2} (K\cdot \frac{\partial }{\partial R_1})-   R_2\cdot \frac{\partial }{\partial R_2} ( K \cdot \frac{\partial }{\partial R_1} )  
\notag\\ &-   R_2 \cdot \frac{\partial }{\partial R_2} (K \cdot \frac{\partial }{\partial R_2}) - f  R_2 \cdot \frac{\partial }{\partial R_2} - K \cdot R_2 ( K \cdot \frac{\partial }{\partial R_1} ) -  K \cdot R_2 ( K \cdot \frac{\partial }{\partial R_2})- f  K \cdot R_2   \notag\\  &- M^2_2  R_2 \cdot \frac{\partial }{\partial R_1}  - R_2 \cdot \frac{\partial }{\partial R_1}(  K \cdot \frac{\partial }{\partial R_2})\} \vec{\alpha} + \frac{1}{2}K^2( R_1 \cdot \frac{\partial }{\partial R_1}+ K \cdot R_1  -  R_2 \cdot \frac{\partial }{\partial R_2}- K \cdot R_2 \notag\\ & + 2  R_1 \cdot \frac{\partial }{\partial R_2}) \vec{\beta}_{23} + \frac{1}{2}K^2( R_1 \cdot \frac{\partial }{\partial R_1}+ K \cdot R_1  -  R_2 \cdot \frac{\partial }{\partial R_2}- K \cdot R_2 - 2 R_2\cdot \frac{\partial }{\partial R_1} )\vec{\beta}_{13} \notag\\ &+ \frac{1}{2} K^2(- R_1\cdot \frac{\partial }{\partial R_1} - K \cdot R_1  +  R_2 \cdot \frac{\partial }{\partial R_2} + K \cdot R_2 ) \vec{\beta}_{12}\,.  \label{2.20} 
\end{align}   
The second pair of vectors is 
\begin{align}
	v_1^\mu = & 0 \,,\notag\\
	v_2^\mu =&\{l^2_2 l_1 \cdot K - l_2 \cdot K l_1 \cdot l_2 - (l_2 \cdot K)^2 +K^2 l^2_2 \} l^\mu_1 \notag\\ &+ (l^2_1 l_2 \cdot K + l_1 \cdot K l_2 \cdot K - l_1 \cdot l_2 l_1 \cdot K  -K^2 l_1 \cdot l_2 )l^\mu_2 \notag\\ &+ \{(l_1 \cdot l_2)^2 +l_1 \cdot l_2 l_2 \cdot K -l^2_1 l^2_2 - l_1 \cdot K l^2_2\}K^\mu  \,.
\end{align}
and the corresponding differential equation is
\begin{align}
	0 =&   \{(D-2) M^2_1  K\cdot \frac{\partial }{\partial R_2} + 2(D-2)K\cdot \frac{\partial }{\partial R_1} ( K\cdot \frac{\partial }{\partial R_2}) + (D-2)K\cdot \frac{\partial }{\partial R_1} ( K\cdot \frac{\partial }{\partial R_1}) \notag\\& + (D-2) f K\cdot \frac{\partial }{\partial R_1} + (D-2) K^2 K\cdot \frac{\partial }{\partial R_1} + (D-2) K^2 K\cdot \frac{\partial }{\partial R_2} + (D-2) f K^2 \notag\\& + M^2_2  R_2\cdot\frac{\partial }{\partial R_1} ( K\cdot \frac{\partial }{\partial R_1} )+ M^2_2  K^2 R_2\cdot\frac{\partial }{\partial R_1} +  R_2\cdot\frac{\partial }{\partial R_1} ( K\cdot \frac{\partial }{\partial R_1} ( K\cdot \frac{\partial }{\partial R_2} )) \notag\\&+ f R_2\cdot\frac{\partial }{\partial R_1} ( K\cdot \frac{\partial }{\partial R_2} )
	+ M^2_1  R_2\cdot\frac{\partial }{\partial R_2} ( K\cdot \frac{\partial }{\partial R_2} )  + 2  R_2\cdot\frac{\partial }{\partial R_2} ( K\cdot \frac{\partial }{\partial R_1} ( K\cdot \frac{\partial }{\partial R_2} )) \notag\\&+  R_2\cdot\frac{\partial }{\partial R_2} ( K\cdot \frac{\partial }{\partial R_1} ( K\cdot \frac{\partial }{\partial R_1} ))  + f  R_2\cdot\frac{\partial }{\partial R_2} ( K\cdot \frac{\partial }{\partial R_1} )  + K^2 R_2\cdot\frac{\partial }{\partial R_2} ( K\cdot \frac{\partial }{\partial R_1} ) \notag\\&+ K^2 R_2\cdot\frac{\partial }{\partial R_2} ( K\cdot \frac{\partial }{\partial R_2} ) + f K^2 R_2\cdot\frac{\partial }{\partial R_2} -  M^2_1  M^2_2 K \cdot R_2  -  M^2_2  K \cdot R_2  K\cdot \frac{\partial }{\partial R_1}  \notag\\&     + K \cdot R_2 K\cdot \frac{\partial }{\partial R_1} ( K\cdot \frac{\partial }{\partial R_2})  + f K \cdot R_2 K\cdot \frac{\partial }{\partial R_2} + K \cdot R_2  K\cdot \frac{\partial }{\partial R_1} ( K\cdot \frac{\partial }{\partial R_1}) \notag\\&+ 2 f K \cdot R_2 K\cdot \frac{\partial }{\partial R_1} + f^2  K \cdot R_2 \} \vec{\alpha} \notag\\      
	&+  \frac{K^2}{2}\{ 2(D-2)K\cdot \frac{\partial }{\partial R_2} + (D-2)K\cdot \frac{\partial }{\partial R_1} + (D-2)K^2  + R_2\cdot\frac{\partial }{\partial R_1} ( K\cdot \frac{\partial }{\partial R_2} )  \notag\\& + 2 R_2\cdot\frac{\partial }{\partial R_2} ( K\cdot \frac{\partial }{\partial R_2} ) + R_2\cdot\frac{\partial }{\partial R_2} ( K\cdot \frac{\partial }{\partial R_1} ) + K^2 R_2\cdot\frac{\partial }{\partial R_2}  - 2 K\cdot R_2 \frac{\partial }{\partial R_2}\cdot \frac{\partial }{\partial R_2} \notag\\&+ K \cdot R_2 K\cdot \frac{\partial }{\partial R_1}  - K \cdot R_2 \frac{\partial }{\partial R_1}\cdot \frac{\partial }{\partial R_2} + f  K \cdot R_2     \}  \vec{\beta}_{23}  \notag\\
	&+  \frac{K^2}{2}\{(D-2)K\cdot \frac{\partial }{\partial R_1} + (D-2)K^2 + R_2\cdot\frac{\partial }{\partial R_1} ( K\cdot \frac{\partial }{\partial R_2} ) + 2 R_2\cdot\frac{\partial }{\partial R_1} ( K\cdot \frac{\partial }{\partial R_1} ) \notag\\&+ 2 K^2 R_2\cdot\frac{\partial }{\partial R_1}      +  R_2\cdot\frac{\partial }{\partial R_2} ( K\cdot \frac{\partial }{\partial R_1} ) + K^2  R_2\cdot\frac{\partial }{\partial R_2} - 2 M^2_1 K \cdot R_2 + f  K \cdot R_2  \notag\\&- K \cdot R_2 \frac{\partial }{\partial R_1}\cdot \frac{\partial }{\partial R_2}   - K \cdot R_2  K\cdot \frac{\partial }{\partial R_1}  \}   \vec{\beta}_{13} \notag\\ 
	&+ \frac{K^2}{2} \{ -(D-2)K\cdot \frac{\partial }{\partial R_1} - (D-2)K^2 - R_2\cdot\frac{\partial }{\partial R_1} ( K\cdot \frac{\partial }{\partial R_2} ) - R_2\cdot\frac{\partial }{\partial R_2} ( K\cdot \frac{\partial }{\partial R_1} ) \notag\\& - K^2 R_2\cdot\frac{\partial }{\partial R_2} - K \cdot R_2 K\cdot \frac{\partial }{\partial R_1}  + K \cdot R_2 \frac{\partial }{\partial R_1}\cdot \frac{\partial }{\partial R_2} - f K \cdot R_2          \}   \vec{\beta}_{12}\,.  \label{2.22}
\end{align}

Having obtained the complete set of differential equations, we will solve the reduction coefficient $\vec{\alpha}$ in the next section.

\section{Solving procedure }\label{sec:solve}

Although we have the complete set of differential equations, finding a closed analytic expression is very difficult in general. Mathematically, there is no general solving algorithm for multivariable partial differential equations. Thus, we set a slightly less ambitious goal: 
		to find the recurrence relations of the expansion coefficients in \eqref{3.1}.  
		
Before presenting the recurrence relation, let us give two remarks:
\begin{itemize}
	\item First the expansion coefficient $\vec{\alpha}_{abnmk}$ in \eqref{3.1} does not correspond to a single tensor integral. Given a tensor 
	integral of the form $\int \frac{d^D l_1}{i \pi^{D/2}}\frac{d^D l_2}{i \pi^{D/2}} \frac{(l_1 \cdot R_1)^A (l_2 \cdot R_2)^B}{D_1 D_2 D_3}$, it will contribute to all coefficients satisfying the relation $a+2 n+k=A$ and $b+2m+k=B$. From this point of view, the recurrence relation is different from
	the familiar IBP relations relating different Feynman integrals. 
	
	\item Secondly, as we have mentioned in the section of introduction, a key point of integrals in a family is that there are some structures, which
	are expressed through the recurrence relations. We can obtain  some kinds of recurrence relations without using the generating function as in \cite{Feng:2021enk,Hu:2021nia}. However, with the fixed tensor rank, solving these relations with constraints is a little bit complicated. This is one of the reasons for introducing the generating function in  \cite{Feng:2022hyg}. Using the generating function, we will be able to get different kinds of recurrence relations, which will be easier to solve.

\end{itemize}

\subsection{The setup}

The key observation is that $\vec{\alpha}$ is a function of  scalars and auxiliary vectors $R_i$ can only appear in the numerator, thus we can write 
\begin{align}
	\vec{\alpha} = \sum_{abnmk} \vec{\alpha}_{abnmk} y^a_1 y^b_2 x^n_{11} x^m_{22} x^k_{12} \label{3.1} \,.
\end{align} 
where $y_i, x_{ij}$ are five possible independent Lorentz contractions of $R_i$. To get simpler differential equations, we choose them as 
\begin{align}
	&y_1 = K \cdot R_1 \,,~~~~ y_2 = K \cdot R_2\,,~~~~ x_{11} = K^2 R^2_1 - ( K \cdot R_1)^2 \,,\notag\\ & x_{22} = K^2 R^2_2 - ( K \cdot R_2)^2 \,,~~  x_{12} = K^2 R_1 \cdot R_2 -  K \cdot R_1 K \cdot R_2 \,.
\end{align}
Similarly, we can expand $\vec{\beta}$ as
\begin{align}
	\vec{\beta}_{ij}=	\sum_{abnmk=0}^{\infty} \vec{\beta}_{ij;abnmk} y^a_1 y^b_2 x^n_{11} x^m_{22} x^k_{12}\,.\label{3.3}
\end{align}
In this paper, by our iterative construction, all coefficients corresponding to lower subsectors, i.e., $\vec{\beta}_{ij}$, are regarded as predetermined values and may consequently be handled as constants. The calculations of these terms are provided in the appendix \ref{sec:Tadpole result}. In the following derivation of the iterative relation for $\vec{\alpha}_{abnmk}$, cases may arise where some indices of  $a,b,n,m,k$ become negative. However, no such terms actually appear in the expansion. We therefore define: $\vec{\alpha}_{abnmk}=0$ whenever any index is negative and the same applies to $\vec{\beta}_{ij}$.

Before going into the details, let us notice that several coefficients are known already. If we set $R_i$ to zero in \eqref{3.1}, only the term $\vec{\alpha} _{00000}$ is left, while \eqref{2.1} tell us it is the basis $I_1$ at this limit, thus we have:
\begin{align}
	\vec{\alpha} _{00000} = (1,0,0,0,0,0,0)  \,.\label{3.4}
\end{align}
Secondly, taking $K\cdot \frac{\partial}{\partial R_1}$ at the both sides of \eqref{2.1} and then set $R_i = 0$, we get the left hand side is $K^2 I_2$ while the only left term at the right handside is $K^2 \vec{\alpha}_{10000}$ thus we can solve
\begin{align}
	\vec{\alpha} _{10000} = (0,1,0,0,0,0,0)\,.
\end{align}
Similarly, using $K\cdot \frac{\partial}{\partial R_2}$ we find
\begin{align}
	\vec{\alpha} _{01000} = (0,0,1,0,0,0,0) \,.
\end{align}
Finally taking both $K\cdot \frac{\partial}{\partial R_1}$ and $K\cdot \frac{\partial}{\partial R_2}$, we find
\begin{align}
	\vec{\alpha} _{11000} = (0,0,0,1,0,0,0)   \,.\label{3.7}
\end{align}
When we solve the differential equations, there are some initial conditions. Thus, \eqref{3.4} to \eqref{3.7} will be these initial conditions. The number of master integrals in the top sector will be the number of necessary initial conditions. Thus, our solving procedure provides another way to determine the number of bases in the top sector.

\subsection{The first group of equations}

Before deriving the recurrence relations for the coefficients, we first re-express the differential operator as follows:
\begin{align}
	\frac{\partial{}}{\partial{R_1}} = K \frac{\partial{}}{\partial{y_1}} + (2 K^2 R_1 - 2(K\cdot R_1 )K)\frac{\partial{}}{\partial{x_{11}}} + ( K^2 R_2 - (K\cdot R_2 )K)\frac{\partial{}}{\partial{x_{12}}} \,,\\ 
	\frac{\partial{}}{\partial{R_2}} = K \frac{\partial{}}{\partial{y_2}} + (2 K^2 R_2 - 2(K\cdot R_2 )K)\frac{\partial{}}{\partial{x_{22}}} + ( K^2 R_1 - (K\cdot R_1 )K)\frac{\partial{}}{\partial{x_{12}}} \,.
\end{align}
Substituting these differential operators into \eqref{2.9}, \eqref{2.10} and \eqref{2.11}, one can see that above three equations become:
\begin{align}
	0 &= -2(D-1)s_0 \frac{\partial \vec{\alpha}}{\partial x_{11}} - s_0 x_{22}\frac{\partial^2 \vec{\alpha}}{\partial x^2_{12}}- 4 s_0 x_{12}\frac{\partial^2 \vec{\alpha}}{\partial x_{11}\partial x_{12}}- 4 s_0 x_{11}\frac{\partial^2 \vec{\alpha}}{\partial x^2_{11}}\nonumber \\ & ~~~~- s_0 \frac{\partial^2 \vec{\alpha}}{\partial y^2_1} + M^2_1 \vec{\alpha} + s_0 \vec{\beta}_{23}\,,\\
		0 &= -2(D-1)s_0 \frac{\partial \vec{\alpha}}{\partial x_{22}} - s_0 x_{11}\frac{\partial^2 \vec{\alpha}}{\partial x^2_{12}}- 4 s_0 x_{12}\frac{\partial^2 \vec{\alpha}}{\partial x_{22}\partial x_{12}}- 4 s_0 x_{22}\frac{\partial^2 \vec{\alpha}}{\partial x^2_{22}}\nonumber \\ & ~~~~- s_0 \frac{\partial^2 \vec{\alpha}}{\partial y^2_2} + M^2_2 \vec{\alpha} + s_0 \vec{\beta}_{13}\,,\\
		0 &= -2(D-1)s_0 \frac{\partial \vec{\alpha}}{\partial x_{12}} - 2 s_0 x_{12}\frac{\partial^2 \vec{\alpha}}{\partial x^2_{12}}-4 s_0 x_{22}\frac{\partial^2 \vec{\alpha}}{\partial x_{22}\partial x_{12}}-4 s_0 x_{11}\frac{\partial^2 \vec{\alpha}}{\partial x_{11}\partial x_{12}}\nonumber \\ & ~~~~- 8 s_0 x_{12}\frac{\partial^2 \vec{\alpha}}{\partial x_{11}\partial x_{22}}  -2 s_0 \frac{\partial \vec{\alpha}}{\partial y_1}-2 s_0 \frac{\partial \vec{\alpha}}{\partial y_2}- 2 s_0 \frac{\partial^2 \vec{\alpha}}{\partial y_1 \partial y_2}\nonumber \\ & ~~~~ - ((M_1^2 +M_2^2 -M_3^2 + s_0)\vec{\alpha} - s_0\vec{\beta}_{12}+ s_0\vec{\beta}_{13}+ s_0\vec{\beta}_{23})\,,
\end{align}
with $s_0 = K^2$. Taking \eqref{3.1}, \eqref{3.3} into the set of differential equations and expanding it, we obtain the following three  relations:
\begin{align}
K^2 \vec{\beta} _{23;a bnmk} =& 2(n+1)s_0(D+ 2k +2n -1) \vec{\alpha} _{a b(n+1)mk}- M^2_1  \vec{\alpha} _{a bnmk}\notag\\ &+ (a+1)(a+2)s_0 \vec{\alpha} _{(a+2) bnmk} + (k+2)(k+1)s_0 \vec{\alpha} _{a bn(m-1)(k+2)}\,, \label{3.13} \\
	K^2 \vec{\beta} _{13;a bnmk} =& 2(m+1)s_0(D+ 2k +2m -1) \vec{\alpha} _{a bn(m+1)k}- M^2_2  \vec{\alpha} _{a bnmk}\notag\\ &+ (b+1)(b+2)s_0 \vec{\alpha} _{a (b+2)nmk} + (k+2)(k+1)s_0 \vec{\alpha} _{a b(n-1)m(k+2)} \,,\label{3.14} \\
\vec{\beta} _{12;a bnmk}-\vec{\beta} &_{13;a bnmk}-\vec{\beta} _{23;a bnmk}= \frac{M^2_1+M^2_2-M^2_3+s_0}{s_0}\vec{\alpha} _{a bnmk} \notag\\
	+2(k+1)(D & +k + 2(m+n)-1)\vec{\alpha} _{a bnm(k+1)} + 8(m+1)(n+1)\vec{\alpha}_{ab(n+1)(m+1)(k-1)} \notag\\
	 + 2(a+1)(b+&1)\vec{\alpha} _{(a+1)(b+1)nmk} + 2(a+1)\vec{\alpha} _{(a+1)bnmk}+2(b+1)\vec{\alpha} _{a(b+1)nmk} \,.\label{3.15}
\end{align}

Now we solve the above three relations. We will show that using \eqref{3.13}, \eqref{3.14} and \eqref{3.15}, we can express any $\vec{\alpha}_{abmnk}$ in terms of $\vec{\alpha}_{a^\prime b^\prime000}$. There are specific steps as follows:\\

{\bf Step One:}  We first solve $\vec{\alpha}_{abnmk}$. Using \eqref{3.15}, after replacing $n$, $m$ and $k$ by $(n-1, m-1, k+1)$, we can obtain the following recurrence relation:
\begin{align}
	&\vec{\alpha} _{a bnmk}= -\frac{1}{8 m n s_0}\{M_1^2 \vec{\alpha} _{a b(n-1)(m-1)(k+1)}+M_2^2 \vec{\alpha} _{a b(n-1)(m-1)(k+1)}-M_3^2 \vec{\alpha} _{a b(n-1)(m-1)(k+1)}\notag\\ & +s_0 \vec{\alpha} _{a b(n-1)(m-1)(k+1)}-16 s_0 \vec{\alpha} _{a
		b(n-1)(m-1)(k+2)}+4 D s_0 \vec{\alpha} _{a b(n-1)(m-1)(k+2)}\notag\\ & -4 k s_0 \vec{\alpha} _{a b(n-1)(m-1)(k+2)}+2 D k s_0 \vec{\alpha} _{a b(n-1)(m-1)(k+2)}+2 k^2 s_0 \vec{\alpha} _{a b(n-1)(m-1)(k+2)}\notag\\ & +8 m s_0 \vec{\alpha}
	_{a b(n-1)(m-1)(k+2)}+4 k m s_0 \vec{\alpha} _{a b(n-1)(m-1)(k+2)}+8 n s_0 \vec{\alpha} _{a b(n-1)(m-1)(k+2)}\notag\\ & +4 k n s_0 \vec{\alpha} _{a b(n-1)(m-1)(k+2)}+2 s_0 \vec{\alpha} _{(a+1) b(n-1)(m-1)(k+1)}+2 a s_0
	\vec{\alpha} _{(a+1) b(n-1)(m-1)(k+1)}\notag\\ & +2 s_0 \vec{\alpha} _{a (b+1)(n-1)(m-1)(k+1)}+2 b s_0 \vec{\alpha} _{a (b+1)(n-1)(m-1)(k+1)}+2 s_0 \vec{\alpha} _{(a+1) (b+1)(n-1)(m-1)(k+1)}\notag\\ & +2 a s_0 \vec{\alpha} _{(a+1)
		(b+1)(n-1)(m-1)(k+1)}+2 b s_0 \vec{\alpha} _{(a+1) (b+1)(n-1)(m-1)(k+1)}-s_0 \vec{\beta} _{12;a b(n-1)(m-1)(k+1)}\notag\\ & +2 a b s_0 \vec{\alpha} _{(a+1) (b+1)(n-1)(m-1)(k+1)}+s_0 \vec{\beta} _{13;a
		b(n-1)(m-1)(k+1)}+s_0 \vec{\beta} _{23;a b(n-1)(m-1)(k+1)}\}\,,
\end{align}
 
The coefficients $n$ and $m$ of the terms on the right-hand side of this equation decrease by one, whereas the coefficients $a,b,k$ may increase. Note that the equation cannot be used for $n,m=0$. Therefore, we can repeatedly apply this relation to express $\vec{\alpha}_{abnmk}$ as a linear combination of a set of $\vec{\alpha}_{a^\prime b^\prime n^\prime 0k^\prime}$ and $\vec{\alpha}_{a^\prime b^\prime0m^\prime k^\prime}$.\\

{\bf Step Two:}  We now solve $\vec{\alpha}_{abn0k}$. By setting $(n-1, 0)$ for $n$ and $m$ in \eqref{3.13}, we obtain the following recurrence relation:
\begin{align}
	\vec{\alpha} _{abn0k}= \frac{M_1^2 \vec{\alpha} _{ab(n-1)0k}+s_0 \left(-\left(2+3 a+a^2\right) \vec{\alpha} _{(a+2)b(n-1)0k}+\vec{\beta} _{23;ab(n-1)0k}\right)}{2 n (D+2 k+2 n-3)
		s_0}\,.
\end{align}
Thus using this equation, we can recursively solve all $\vec{\alpha} _{abn0k}$ as the function of $\vec{\alpha} _{a^\prime b^\prime 00k^\prime}$ only.\\

{\bf Step Three:}  We now solve $\vec{\alpha}_{ab0mk}$. By setting $(0, m-1)$ for $n$ and $m$ in \eqref{3.14}, we obtain the following recurrence relation:
\begin{align}
	\vec{\alpha} _{ab0mk}= \frac{M_2^2 \vec{\alpha} _{ab0(m-1)k}+s_0 \left(-\left(2+3 b+b^2\right) \vec{\alpha} _{a(b+2)0(m-1)k}+\vec{\beta} _{13;ab0(m-1)k}\right)}{2 m (D+2 k+2 m-3)
		s_0}\,.
\end{align}
Using this equation, we can recursively solve all $\vec{\alpha} _{ab0mk}$ as the function of $\vec{\alpha} _{a^\prime b^\prime00k^\prime}$ only.\\

{\bf Step Four:}  We now solve $\vec{\alpha}_{ab00k}$: the most convenient iterative relation is given by \eqref{3.15}. By setting $n,m=0$ in \eqref{3.15}, we obtain the recurrence relation for $\vec{\alpha}_{ab00k}$, but this equation contains the term $\vec{\alpha}_{ab11(k-1)}$:
\begin{align}
	&\vec{\alpha} _{ab00(k+1)}=\frac{1}{2(D s_0 - s_0 + D k s_0 + k^2 s_0)}\{- M_1^2 \vec{\alpha} _{ab00k} - M_2^2 \vec{\alpha} _{ab00k}+ M_3^2 \vec{\alpha} _{ab00k} - s_0 \vec{\alpha} _{ab00k} \notag\\ & - 8s_0 \vec{\alpha} _{ab11(k-1)}  - 2 s_0 \vec{\alpha} _{a(b+1)00k} - 2 b s_0 \vec{\alpha} _{a(b+1)00k}- 2  s_0 \vec{\alpha} _{(a+1)b00k} - 2 a s_0 \vec{\alpha} _{(a+1)b00k}  \notag\\ & - 2  s_0 \vec{\alpha} _{(a+1)(b+1)00k}- 2 a s_0 \vec{\alpha} _{(a+1)(b+1)00k} - 2b  s_0 \vec{\alpha} _{(a+1)(b+1)00k} - 2 ab s_0 \vec{\alpha} _{(a+1)(b+1)00k} \notag\\ &  + s_0 \vec{\beta} _{12;ab00k}-s_0 \vec{\beta} _{13;ab00k}-s_0 \vec{\beta} _{23;ab00k} \} \,.\label{3.19}
\end{align}

We now solve for $\vec{\alpha}_{ab11k}$.  We take $(0, 0), (0, 1)$ for $n$ and $m$ in equation \eqref{3.13}, and $ (a, 0, 0), (a+2, 0, 0)$ for $a$, $n$ and $m$ in equation \eqref{3.14}, resulting in a total of four equations which contain the following four terms: $\vec{\alpha}_{ab11k}$, $\vec{\alpha}_{ab01k}$, $\vec{\alpha}_{ab10k}$, $\vec{\alpha}_{(a+2)b01k}$. The remaining terms are all of the form $\vec{\alpha}_{a^\prime b^\prime00k^\prime}$. We can explicitly solve these four terms to obtain an expression for $\vec{\alpha}_{ab11k}$ in terms of $\vec{\alpha}_{a^\prime b^\prime00k^\prime}$. 

Thus, after substituting the expression of $\vec{\alpha}_{ab11k}$ into \eqref{3.19} and adjusting the coefficient of $k$, we obtain the solution for $\vec{\alpha}_{ab00k}$:
\begin{align}
	&\vec{\alpha} _{ab00k}= \frac{1}{2 k (D+2 k-5) \left(12+D^2-11 k+2 k^2+D (3 k-7)\right) s_0^2}\{-2 M_1^2 M_2^2 \vec{\alpha} _{ab00(k-2)}\notag\\ & -(D+2 k-5)^2 s_0 \left(M_1^2+M_2^2-M_3^2+s_0\right) \vec{\alpha} _{ab00(k-1)}-2 (1+b) (D+2 k-5)^2 s_0^2 \vec{\alpha}
	_{a(b+1)00(k-1)}\notag\\ & +4 M_1^2 s_0 \vec{\alpha} _{a(b+2)00(k-2)}+6 b M_1^2 s_0 \vec{\alpha} _{a(b+2)00(k-2)}+2 b^2 M_1^2 s_0 \vec{\alpha} _{a(b+2)00(k-2)}\notag\\ & -2 (1+a) (D+2 k-5)^2 s_0^2 \vec{\alpha}
	_{(a+1)b00(k-1)}-2 (1+a) (1+b) (D+2 k-5)^2 s_0^2 \vec{\alpha} _{(a+1)(b+1)00(k-1)} \notag\\ & +4 M_2^2 s_0 \vec{\alpha} _{(a+2)b00(k-2)}+6 a M_2^2 s_0 \vec{\alpha} _{(a+2)b00(k-2)}+2 a^2 M_2^2 s_0 \vec{\alpha}
	_{(a+2)b00(k-2)} \notag\\ & -8 s_0^2 \vec{\alpha} _{(a+2)(b+2)00(k-2)}-12 a s_0^2 \vec{\alpha} _{(a+2)(b+2)00(k-2)}-4 a^2 s_0^2 \vec{\alpha} _{(a+2)(b+2)00(k-2)}\notag\\ & -12 b s_0^2 \vec{\alpha} _{(a+2)(b+2)00(k-2)}-18 a b s_0^2
	\vec{\alpha} _{(a+2)(b+2)00(k-2)}-6 a^2 b s_0^2 \vec{\alpha} _{(a+2)(b+2)00(k-2)} \notag\\ & -4 b^2 s_0^2 \vec{\alpha} _{(a+2)(b+2)00(k-2)}-6 a b^2 s_0^2 \vec{\alpha} _{(a+2)(b+2)00(k-2)}-2 a^2 b^2 s_0^2 \vec{\alpha}
	_{(a+2)(b+2)00(k-2)}\notag\\ & -2 M_1^2 s_0 \vec{\beta} _{13;ab00(k-2)}+4 s_0^2 \vec{\beta} _{13;(a+2)b00(k-2)}+6 a s_0^2 \vec{\beta} _{13;(a+2)b00(k-2)}+2 a^2 s_0^2 \vec{\beta} _{13;(a+2)b00(k-2)} \notag\\ & +(-5+D+2
	k)^2 s_0^2 \left(\vec{\beta} _{12;ab00(k-1)}-\vec{\beta} _{13;ab00(k-1)}-\vec{\beta} _{23;ab00(k-1)}\right)+4 s_0^2 \vec{\beta} _{23;ab01(k-2)} \notag\\ & -4 D s_0^2 \vec{\beta} _{23;ab01(k-2)}-8 (k-2)
	s_0^2 \vec{\beta} _{23;ab01(k-2)}\}\,.
\end{align}

\subsection{The second group of equations}

As shown in the previous subsection, using the first group of relations, we can reduce all $\vec{\alpha}_{abnmk}$ to the linear combinations of a set of $\vec{\alpha}_{a^\prime b^\prime000}$. Consequently, to finish the task, we should use
the remaining two equations to solve $\vec{\alpha}_{ab000}$. Since our focus is $\vec{\alpha}_{ab000}$, when we use 
these two differential equations, we can set $n,m,k=0$ and the equations will be substantially simplified.\\

{\bf Step Five:} Apply differential operators to Equation \eqref{2.20}.  The computation result is:
\begin{align}
0 = & -\frac{1}{2} M_1^2  \vec{\alpha} _{a (b-1)000}-\frac{1}{2} M_2^2  \vec{\alpha} _{a (b-1)000}+\frac{1}{2} M_3^2  \vec{\alpha} _{a (b-1)000}-\frac{1}{2} s_0  \vec{\alpha} _{a (b-1)000}  -M_2^2  \vec{\alpha}_{(a+1) (b-1)000}\notag\\ & -a M_2^2  \vec{\alpha} _{(a+1) (b-1)000}-s_0  \vec{\alpha} _{(a+1) (b-1)000}-a s_0  \vec{\alpha} _{(a+1) (b-1)000}  +\frac{1}{2} M_1^2  \vec{\alpha} _{(a-1) b000} \notag\\ & +\frac{1}{2}
M_2^2  \vec{\alpha} _{(a-1) b000}-\frac{1}{2} M_3^2  \vec{\alpha} _{(a-1) b000}+\frac{1}{2} s_0  \vec{\alpha} _{(a-1) b000}  +\frac{1}{2} a M_1^2  \vec{\alpha} _{a b000}  -\frac{1}{2} b M_1^2  \vec{\alpha}_{a b000}\notag\\ & +\frac{1}{2} a M_2^2  \vec{\alpha} _{a b000}-\frac{1}{2} b M_2^2  \vec{\alpha} _{a b000}-\frac{1}{2} a M_3^2  \vec{\alpha} _{a b000}  +\frac{1}{2} b M_3^2  \vec{\alpha} _{a
	b000}+\frac{3}{2} a s_0  \vec{\alpha} _{a b000} \notag\\ & -\frac{3}{2} b s_0  \vec{\alpha} _{a b000}+a s_0  \vec{\alpha} _{(a+1) b000}+a^2 s_0  \vec{\alpha} _{(a+1) b000}-2 b s_0  \vec{\alpha} _{(a+1) b000}-2
a b s_0  \vec{\alpha} _{(a+1) b000} \notag\\ & +M_1^2  \vec{\alpha} _{(a-1) (b+1)000}+b M_1^2  \vec{\alpha} _{(a-1) (b+1)000}+s_0  \vec{\alpha} _{(a-1) (b+1)000}+b s_0  \vec{\alpha} _{(a-1) (b+1)000} \notag\\ & +2 a s_0
 \vec{\alpha} _{a (b+1)000}-b s_0  \vec{\alpha} _{a (b+1)000}+2 a b s_0  \vec{\alpha} _{a (b+1)000}-b^2 s_0  \vec{\alpha} _{a (b+1)000}-\frac{1}{2} s_0 \vec{\beta} _{23;a (b-1)000} \notag\\ & +\frac{1}{2} s_0 \vec{\beta}_{23;(a-1) b000}+\frac{1}{2} a s_0 \vec{\beta} _{23;a b000}-\frac{1}{2} b s_0 \vec{\beta} _{23;a b000}+s_0 \vec{\beta} _{23;(a-1) (b+1)000}\notag\\ & +b s_0 \vec{\beta} _{23;(a-1) (b+1)000}   -\frac{1}{2} s_0 \vec{\beta} _{13;a
	(b-1)000}  -s_0 \vec{\beta} _{13;(a+1) (b-1)000}-a s_0 \vec{\beta} _{13;(a+1) (b-1)000}\notag\\ & +\frac{1}{2} s_0 \vec{\beta} _{13;(a-1) b000}  +\frac{1}{2} a s_0 \vec{\beta} _{13;a b000}  -\frac{1}{2} b s_0 \vec{\beta} _{13;a
	b000}+\frac{1}{2} s_0 \vec{\beta} _{12;a (b-1)000}  \notag\\ &-\frac{1}{2} s_0 \vec{\beta} _{12;(a-1) b000} -\frac{1}{2} a s_0 \vec{\beta} _{12;a b000}+\frac{1}{2} b s_0 \vec{\beta} _{12;a b000} \,.  \label{3.20}
\end{align}
The same method can be applied to simplify \eqref{2.22}. After replacing $a$ by $a-1$, we get:
\begin{align}
	0 =	& 2 f^2 \vec{\alpha} _{(a-1) (b-1)000}-2 M_1^2 M_2^2 \vec{\alpha} _{(a-1) (b-1)000}+4 a f s_0 \vec{\alpha} _{a (b-1)000}-2 (a+1) M_2^2 s_0 \vec{\alpha} _{(a+1) (b-1)000} \notag\\ & +2 (a+1)^2 M_2^2 s_0 \vec{\alpha} _{(a+1)
		(b-1)000}-2 (a+1) s_0^2 \vec{\alpha} _{(a+1) (b-1)000}+2 (a+1)^2 s_0^2 \vec{\alpha} _{(a+1) (b-1)000}\notag\\ &-4 f s_0 \vec{\alpha} _{(a-1) b000}+4 b f s_0 \vec{\alpha} _{(a-1) b000}+2 D f s_0 \vec{\alpha} _{(a-1)
		b000}-4 a f s_0 \vec{\alpha} _{a b000}+4 a b f s_0 \vec{\alpha} _{a b000} \notag\\ & +2 a D f s_0 \vec{\alpha} _{a b000}-4 a s_0^2 \vec{\alpha} _{a b000}+4 a b s_0^2 \vec{\alpha} _{a b000}+2 a D s_0^2 \vec{\alpha} _{a b000}+4 (a+1)
	s_0^2 \vec{\alpha} _{(a+1) b000} \notag\\ & -4 (a+1)^2 s_0^2 \vec{\alpha} _{(a+1) b000}-4 (a+1) b s_0^2 \vec{\alpha} _{(a+1) b000}+4 (a+1)^2 b s_0^2 \vec{\alpha} _{(a+1) b000}\notag\\ & -2 (a+1) D s_0^2 \vec{\alpha} _{(a+1) b000}+2 (a+1)^2
	D s_0^2 \vec{\alpha} _{(a+1) b000}-6 (b+1) M_1^2 s_0 \vec{\alpha} _{(a-1) (b+1)000} \notag\\ & +2 (b+1)^2 M_1^2 s_0 \vec{\alpha} _{(a-1) (b+1)000}+2 (b+1) D M_1^2 s_0 \vec{\alpha} _{(a-1) (b+1)000}-6 (b+1) s_0^2 \vec{\alpha}
	_{(a-1) (b+1)000} \notag\\ & +2 (b+1)^2 s_0^2 \vec{\alpha} _{(a-1) (b+1)000}+2 (b+1) D s_0^2 \vec{\alpha} _{(a-1) (b+1)000}-12 a (b+1) s_0^2 \vec{\alpha} _{a (b+1)000} \notag\\ & +4 a (b+1)^2 s_0^2 \vec{\alpha} _{a (b+1)000}+4 a
	(b+1) D s_0^2 \vec{\alpha} _{a (b+1)000}-f s_0\vec{\beta} _{12;(a-1) (b-1)000}-s_0^2 \vec{\beta} _{12;(a-1) (b-1)001} \notag\\ & +D s_0^2 \vec{\beta} _{12;(a-1) (b-1)001}-a s_0^2 \vec{\beta} _{12;a (b-1)000}+2
	s_0^2 \vec{\beta} _{12;(a-1) b000}-b s_0^2 \vec{\beta} _{12;(a-1) b000} \notag\\ & -D s_0^2 \vec{\beta} _{12;(a-1) b000}+2 a s_0^2 \vec{\beta} _{12;a b000}-a b s_0^2 \vec{\beta} _{12;a b000}-a D s_0^2 \vec{\beta}
	_{12;a b000}+f s_0 \vec{\beta} _{13;(a-1) (b-1)000} \notag\\ & -2 M_1^2 s_0 \vec{\beta} _{13;(a-1) (b-1)000}+s_0^2 \vec{\beta} _{13;(a-1) (b-1)001}-D s_0^2 \vec{\beta} _{13;(a-1) (b-1)001}+a s_0^2 \vec{\beta}
	_{13;a (b-1)000} \notag\\ & -2 (a+1) s_0^2 \vec{\beta} _{13;(a+1) (b-1)000}+2 (a+1)^2 s_0^2 \vec{\beta} _{13;(a+1) (b-1)000}-2 s_0^2 \vec{\beta} _{13;(a-1) b000} \notag\\ & +b s_0^2 \vec{\beta} _{13;(a-1) b000}+D
	s_0^2 \vec{\beta} _{13;(a-1) b000}-2 a s_0^2 \vec{\beta} _{13;a b000}+a b s_0^2 \vec{\beta} _{13;a b000}+a D s_0^2 \vec{\beta} _{13;a b000} \notag\\ & +f s_0 \vec{\beta} _{23;(a-1) (b-1)000}+s_0^2 \vec{\beta}
	_{23;(a-1) (b-1)001}-D s_0^2 \vec{\beta} _{23;(a-1) (b-1)001}+4 s_0^2 \vec{\beta} _{23;(a-1) (b-1)010} \notag\\ & -4 D s_0^2 \vec{\beta} _{23;(a-1) (b-1)010}+a s_0^2 \vec{\beta} _{23;a (b-1)000}-2
	s_0^2 \vec{\beta} _{23;(a-1) b000}+b s_0^2 \vec{\beta} _{23;(a-1) b000}\notag\\ & +D s_0^2 \vec{\beta} _{23;(a-1) b000}-2 a s_0^2 \vec{\beta} _{23;a b000}+a b s_0^2 \vec{\beta} _{23;a b000}+a D s_0^2 \vec{\beta}
	_{23;a b000} \notag\\ & -4 (b+1) s_0^2 \vec{\beta} _{23;(a-1) (b+1)000}+2 (b+1) D s_0^2 \vec{\beta} _{23;(a-1) (b+1)000} \,.\label{3.21}
\end{align}

Now we want to solve $\vec{\alpha} _{ab000}$: We perform a linear combination of \eqref{3.20} and \eqref{3.21} to eliminate the $\vec{\alpha} _{(a+1) (b-1)000}$-terms. After replacing $a$ by $a-1$, we obtain:
\begin{align}
	&\vec{\alpha} _{a b000} = -\frac{1}{4 (a-1) a (-3+a+D) s_0^2}\{M_1^4 \vec{\alpha} _{(a-2) (b-1)000}-2 M_1^2 M_2^2 \vec{\alpha} _{(a-2) (b-1)000}\notag\\ & +M_2^4 \vec{\alpha} _{(a-2) (b-1)000}-2 M_1^2 M_3^2 \vec{\alpha} _{(a-2) (b-1)000}-2 M_2^2 M_3^2
	\vec{\alpha} _{(a-2) (b-1)000}+M_3^4 \vec{\alpha} _{(a-2) (b-1)000} \notag\\ & +2 M_1^2 s_0 \vec{\alpha} _{(a-2) (b-1)000}+2 M_2^2 s_0 \vec{\alpha} _{(a-2) (b-1)000}-2 M_3^2 s_0 \vec{\alpha} _{(a-2) (b-1)000}+s_0^2
	\vec{\alpha} _{(a-2) (b-1)000} \notag\\ & +2 (a-1) M_1^2 s_0 \vec{\alpha} _{(a-1) (b-1)000}+2 (a-1) M_2^2 s_0 \vec{\alpha} _{(a-1) (b-1)000}-2 (a-1) M_3^2 s_0 \vec{\alpha} _{(a-1) (b-1)000} \notag\\ & +2 (a-1) s_0^2
	\vec{\alpha} _{(a-1) (b-1)000}-4 M_1^2 s_0 \vec{\alpha} _{(a-2) b000}+2 (a-1) M_1^2 s_0 \vec{\alpha} _{(a-2) b000}+4 b M_1^2 s_0 \vec{\alpha} _{(a-2) b000}\notag\\ & +2 D M_1^2 s_0 \vec{\alpha} _{(a-2) b000}-4 M_2^2 s_0
	\vec{\alpha} _{(a-2) b000}+2 (a-1) M_2^2 s_0 \vec{\alpha} _{(a-2) b000}+4 b M_2^2 s_0 \vec{\alpha} _{(a-2) b000} \notag\\ & +2 D M_2^2 s_0 \vec{\alpha} _{(a-2) b000}+4 M_3^2 s_0 \vec{\alpha} _{(a-2) b000}-2 (a-1) M_3^2
	s_0 \vec{\alpha} _{(a-2) b000}-4 b M_3^2 s_0 \vec{\alpha} _{(a-2) b000} \notag\\ & -2 D M_3^2 s_0 \vec{\alpha} _{(a-2) b000}-4 s_0^2 \vec{\alpha} _{(a-2) b000}+2 (a-1) s_0^2 \vec{\alpha} _{(a-2) b000}+4 b s_0^2 \vec{\alpha}
	_{(a-2) b000}+2 D s_0^2 \vec{\alpha} _{(a-2) b000} \notag\\ & -4 (a-1) M_1^2 s_0 \vec{\alpha} _{(a-1) b000}+2 (a-1)^2 M_1^2 s_0 \vec{\alpha} _{(a-1) b000}+2 (a-1) b M_1^2 s_0 \vec{\alpha} _{(a-1) b000} \notag\\ & +2 (a-1) D
	M_1^2 s_0 \vec{\alpha} _{(a-1) b000}-4 (a-1) M_2^2 s_0 \vec{\alpha} _{(a-1) b000}+2 (a-1)^2 M_2^2 s_0 \vec{\alpha} _{(a-1) b000} \notag\\ & +2 (a-1) b M_2^2 s_0 \vec{\alpha} _{(a-1) b000}+2 (a-1) D M_2^2 s_0
	\vec{\alpha} _{(a-1) b000}+4 (a-1) M_3^2 s_0 \vec{\alpha} _{(a-1) b000} \notag\\ & -2 (a-1)^2 M_3^2 s_0 \vec{\alpha} _{(a-1) b000}-2 (a-1) b M_3^2 s_0 \vec{\alpha} _{(a-1) b000}-2 (a-1) D M_3^2 s_0 \vec{\alpha} _{(a-1)
		b000} \notag\\ & -12 (a-1) s_0^2 \vec{\alpha} _{(a-1) b000}+6 (a-1)^2 s_0^2 \vec{\alpha} _{(a-1) b000}+6 (a-1) b s_0^2 \vec{\alpha} _{(a-1) b000} \notag\\ & +6 (a-1) D s_0^2 \vec{\alpha} _{(a-1) b000}-8 M_1^2 s_0 \vec{\alpha}
	_{(a-2) (b+1)000}+4 (a-1) M_1^2 s_0 \vec{\alpha} _{(a-2) (b+1)000} \notag\\ & -4 b M_1^2 s_0 \vec{\alpha} _{(a-2) (b+1)000}+4 (a-1) b M_1^2 s_0 \vec{\alpha} _{(a-2) (b+1)000}+4 b^2 M_1^2 s_0 \vec{\alpha} _{(a-2)
		(b+1)000} \notag\\ & +4 D M_1^2 s_0 \vec{\alpha} _{(a-2) (b+1)000}+4 b D M_1^2 s_0 \vec{\alpha} _{(a-2) (b+1)000}-8 s_0^2 \vec{\alpha} _{(a-2) (b+1)000} \notag\\ & +4 (a-1) s_0^2 \vec{\alpha} _{(a-2) (b+1)000}-4 b s_0^2 \vec{\alpha}
	_{(a-2) (b+1)000}+4 (a-1) b s_0^2 \vec{\alpha} _{(a-2) (b+1)000} \notag\\ & +4 b^2 s_0^2 \vec{\alpha} _{(a-2) (b+1)000}+4 D s_0^2 \vec{\alpha} _{(a-2) (b+1)000}+4 b D s_0^2 \vec{\alpha} _{(a-2) (b+1)000}-16 (a-1)
	s_0^2 \vec{\alpha} _{(a-1) (b+1)000} \notag\\ & +8 (a-1)^2 s_0^2 \vec{\alpha} _{(a-1) (b+1)000}-12 (a-1) b s_0^2 \vec{\alpha} _{(a-1) (b+1)000}+8 (a-1)^2 b s_0^2 \vec{\alpha} _{(a-1) (b+1)000} \notag\\ & +4 (a-1) b^2 s_0^2
	\vec{\alpha} _{(a-1) (b+1)000}+8 (a-1) D s_0^2 \vec{\alpha} _{(a-1) (b+1)000}+8 (a-1) b D s_0^2 \vec{\alpha} _{(a-1) (b+1)000} \notag\\ & -M_1^2 s_0 \vec{\beta} _{12;(a-2) (b-1)000}-M_2^2 s_0 \vec{\beta} _{12;(a-2)
		(b-1)000}+M_3^2 s_0 \vec{\beta} _{12;(a-2) (b-1)000}-s_0^2 \vec{\beta} _{12;(a-2) (b-1)000} \notag\\ & -2 s_0^2 \vec{\beta} _{12;(a-2) (b-1)001}+2 D s_0^2 \vec{\beta} _{12;(a-2) (b-1)001}+4 s_0^2 \vec{\beta}
	_{12;(a-2) b000}-2 (a-1) s_0^2 \vec{\beta} _{12;(a-2) b000}\notag\\ & -2 b s_0^2 \vec{\beta} _{12;(a-2) b000}-2 D s_0^2 \vec{\beta} _{12;(a-2) b000}+4 (a-1) s_0^2 \vec{\beta} _{12;(a-1) b000}-2
	(a-1)^2 s_0^2 \vec{\beta} _{12;(a-1) b000} \notag\\ & -2 (a-1) D s_0^2 \vec{\beta} _{12;(a-1) b000}-3 M_1^2 s_0 \vec{\beta} _{13;(a-2) (b-1)000}+M_2^2 s_0 \vec{\beta} _{13;(a-2) (b-1)000}\notag\\ & -M_3^2 s_0 \vec{\beta}
	_{13;(a-2) (b-1)000}  +s_0^2 \vec{\beta} _{13;(a-2) (b-1)000}+2 s_0^2 \vec{\beta} _{13;(a-2) (b-1)001}-2 D s_0^2 \vec{\beta} _{13;(a-2) (b-1)001} \notag\\ & -4 s_0^2 \vec{\beta} _{13;(a-2) b000} +2
	(a-1) s_0^2 \vec{\beta} _{13;(a-2) b000}+2 b s_0^2 \vec{\beta} _{13;(a-2) b000}+2 D s_0^2 \vec{\beta} _{13;(a-2) b000}\notag\\ & -4 (a-1) s_0^2 \vec{\beta} _{13;(a-1) b000} +2 (a-1)^2 s_0^2 \vec{\beta}
	_{13;(a-1) b000}+2 (a-1) D s_0^2 \vec{\beta} _{13;(a-1) b000} \notag\\ & +M_1^2 s_0 \vec{\beta} _{23;(a-2) (b-1)000} +M_2^2 s_0 \vec{\beta} _{23;(a-2) (b-1)000}-M_3^2 s_0 \vec{\beta} _{23;(a-2)
		(b-1)000}+s_0^2 \vec{\beta} _{23;(a-2) (b-1)000}\notag\\ & +2 s_0^2 \vec{\beta} _{23;(a-2) (b-1)001} -2 D s_0^2 \vec{\beta} _{23;(a-2) (b-1)001}+8 s_0^2 \vec{\beta} _{23;(a-2) (b-1)010}-8 D s_0^2
	\vec{\beta} _{23;(a-2) (b-1)010}\notag\\ & -4 s_0^2 \vec{\beta} _{23;(a-2) b000} +2 (a-1) s_0^2 \vec{\beta} _{23;(a-2) b000}+2 b s_0^2 \vec{\beta} _{23;(a-2) b000}+2 D s_0^2 \vec{\beta} _{23;(a-2) b000} \notag\\ & -4
	(a-1) s_0^2 \vec{\beta} _{23;(a-1) b000}  +2 (a-1)^2 s_0^2 \vec{\beta} _{23;(a-1) b000}+2 (a-1) D s_0^2 \vec{\beta} _{23;(a-1) b000} \notag\\ & -8 s_0^2 \vec{\beta} _{23;(a-2) (b+1)000} +4 (a-1) s_0^2 \vec{\beta}
	_{23;(a-2) (b+1)000}-8 b s_0^2 \vec{\beta} _{23;(a-2) (b+1)000} \notag\\ & +4 (a-1) b s_0^2 \vec{\beta} _{23;(a-2) (b+1)000} +4 D s_0^2 \vec{\beta} _{23;(a-2) (b+1)000}+4 b D s_0^2 \vec{\beta} _{23;(a-2)
		(b+1)000}\}\,.
\end{align}
Note that this recurrence relation cannot be used for $a=-1$ and 0,  so it can express all $\vec{\alpha} _{a b000}$ in terms of $\vec{\alpha} _{1 b^\prime000}$ and $\vec{\alpha} _{0 b^\prime000}$.\\

{\bf Step Six:} The final step is to solve for $\vec{\alpha} _{1 b000}$ and $\vec{\alpha} _{0 b000}$. The solution simply involves solving a system of linear equations. We take $(1, b-2)$, $(1, b-1)$, $(0, b-1)$ for $a$ and $b$ in equation \eqref{3.20}, and $ (1, b-2)$, $(1, b-1)$ for $a$ and $b$ in equation \eqref{3.21}, resulting in a total of five equations. We treat these terms as unknown variables, which are   $\vec{\alpha} _{1 b000}$, $\vec{\alpha} _{0 b000}$, $\vec{\alpha} _{2 (b-3)000}$, $\vec{\alpha} _{2 (b-2)000}$ and $\vec{\alpha} _{2 (b-1)000}$. Then these five equations can be considered linearly independent, with the first two solutions yielding our desired recurrence relations:
\begin{align}
	&\vec{\alpha} _{0b000} = -\frac{1}{2 (b-1) b s_0}\{-M_3^2 \vec{\alpha} _{0(b-2)000}+s_0 \vec{\alpha} _{0(b-2)000}+M_3^2 \vec{\alpha} _{0(b-1)000}-b M_3^2 \vec{\alpha} _{0(b-1)000} \notag\\ & -3 s_0 \vec{\alpha} _{0(b-1)000}+3 b s_0 \vec{\alpha} _{0(b-1)000}+M_1^2
	(\vec{\alpha} _{0(b-2)000}+(b-1) \vec{\alpha} _{0(b-1)000})  +2 s_0 \vec{\alpha} _{1(b-2)000}\notag\\ & +M_2^2 (\vec{\alpha} _{0(b-2)000}+(b-1) \vec{\alpha} _{0(b-1)000}+2 \vec{\alpha} _{1(b-2)000})-4 s_0 \vec{\alpha}
	_{1(b-1)000}+4 b s_0 \vec{\alpha} _{1(b-1)000} \notag\\ & -s_0 \vec{\beta} _{12;0(b-2)000}+s_0 \vec{\beta} _{12;0(b-1)000}-b s_0 \vec{\beta} _{12;0(b-1)000}+s_0 \vec{\beta} _{13;0(b-2)000}-s_0 \vec{\beta} _{13;0(b-1)000} \notag\\ & +b s_0 \vec{\beta}
	_{13;0(b-1)000}  +2 s_0 \vec{\beta} _{13;1(b-2)000}+s_0 \vec{\beta} _{23;0(b-2)000}-s_0 \vec{\beta} _{23;0(b-1)000}+b s_0 \vec{\beta} _{23;0(b-1)000}\}\,,
\end{align}
\begin{align}
&\vec{\alpha} _{1b000}  =  \frac{1}{4 (b-1) b (-6+2 b+3 D)
	s_0^3} \{-M_2^6 \vec{\alpha} _{0(b-3)000}-M_1^4 (M_2^2 \vec{\alpha} _{0(b-3)000}+s_0 (\vec{\alpha} _{0(b-3)000} \notag\\ & +(7-2 b-4 D) \vec{\alpha} _{0(b-2)000}-2 (b-1) (-5+2 b+2 D) \vec{\alpha}
	_{0(b-1)000}))+M_2^4 (2 M_3^2 \vec{\alpha} _{0(b-3)000} \notag\\ & -s_0 (3 \vec{\alpha} _{0(b-3)000}+(-13+6 b+2 D) \vec{\alpha} _{0(b-2)000}+2 \vec{\alpha} _{1(b-3)000}-6 \vec{\alpha} _{1(b-2)000}+2 b \vec{\alpha}
	_{1(b-2)000} \notag\\ & +2 D \vec{\alpha} _{1(b-2)000}-\vec{\beta} _{12;0(b-3)000}+\vec{\beta} _{13;0(b-3)000}+\vec{\beta} _{23;0(b-3)000}))+M_1^2 (2 M_2^4 \vec{\alpha} _{0(b-3)000} \notag\\ & +M_2^2 (2
	M_3^2 \vec{\alpha} _{0(b-3)000}+s_0 (2 (-3+2 b+D) \vec{\alpha} _{0(b-2)000}+2 (b-1) \vec{\alpha} _{0(b-1)000}-2 \vec{\alpha} _{1(b-3)000}\notag\\ & -14 \vec{\alpha} _{1(b-2)000}+6 b \vec{\alpha} _{1(b-2)000}+6 D \vec{\alpha}
	_{1(b-2)000}+\vec{\beta} _{12;0(b-3)000}+3 \vec{\beta} _{13;0(b-3)000}-\vec{\beta} _{23;0(b-3)000})) \notag\\ & +s_0 (2 M_3^2 (\vec{\alpha} _{0(b-3)000}-2 (-1+D) \vec{\alpha}
	_{0(b-2)000}-(b-1) (-5+2 b+2 D) \vec{\alpha} _{0(b-1)000}) \notag\\ &+s_0 (-2 \vec{\alpha} _{0(b-3)000}+(-4+6 D) \vec{\alpha} _{0(b-2)000}+12 \vec{\alpha} _{0(b-1)000}-16 b \vec{\alpha} _{0(b-1)000}+4 b^2 \vec{\alpha}
	_{0(b-1)000}\notag\\ & -8 D \vec{\alpha} _{0(b-1)000}+8 b D \vec{\alpha} _{0(b-1)000}-2 \vec{\alpha} _{1(b-3)000}-10 \vec{\alpha} _{1(b-2)000}+2 b \vec{\alpha} _{1(b-2)000} +8 D \vec{\alpha} _{1(b-2)000}\notag\\ & +32 \vec{\alpha} _{1(b-1)000}-44 b
	\vec{\alpha} _{1(b-1)000}+12 b^2 \vec{\alpha} _{1(b-1)000}-14 D \vec{\alpha} _{1(b-1)000}+14 b D \vec{\alpha} _{1(b-1)000} \notag\\ & +\vec{\beta} _{12;0(b-3)000}+7 \vec{\beta} _{12;0(b-2)000}-2 b \vec{\beta} _{12;0(b-2)000}-4
	D \vec{\beta} _{12;0(b-2)000}-10 \vec{\beta} _{12;0(b-1)000} \notag\\ & +14 b \vec{\beta} _{12;0(b-1)000}-4 b^2 \vec{\beta} _{12;0(b-1)000}+4 D \vec{\beta} _{12;0(b-1)000}-4 b D \vec{\beta} _{12;0(b-1)000}+3 \vec{\beta} _{13;0(b-3)000} \notag\\ & -19 \vec{\beta} _{13;0(b-2)000}+10 b \vec{\beta} _{13;0(b-2)000}+4 D \vec{\beta} _{13;0(b-2)000}+10 \vec{\beta} _{13;0(b-1)000}-14 b \vec{\beta} _{13;0(b-1)000} \notag\\ & +4 b^2 \vec{\beta} _{13;0(b-1)000}-4 D \vec{\beta} _{13;0(b-1)000}+4 b D \vec{\beta} _{13;0(b-1)000}-20 \vec{\beta} _{13;1(b-2)000}+8 b \vec{\beta} _{13;1(b-2)000} \notag\\ & +8 D
	\vec{\beta} _{13;1(b-2)000}-\vec{\beta} _{23;0(b-3)000}-7 \vec{\beta} _{23;0(b-2)000}+2 b \vec{\beta} _{23;0(b-2)000}+4 D \vec{\beta} _{23;0(b-2)000} \notag\\ & +10 \vec{\beta} _{23;0(b-1)000}-14 b \vec{\beta} _{23;0(b-1)000}+4 b^2 \vec{\beta} _{23;0(b-1)000}-4 D \vec{\beta} _{23;0(b-1)000}+4 b D \vec{\beta} _{23;0(b-1)000}))) \notag\\ & +M_2^2 (-M_3^4 \vec{\alpha} _{0(b-3)000}+M_3^2
	s_0 (4 \vec{\alpha} _{0(b-3)000}+2 (-8+4 b+D) \vec{\alpha} _{0(b-2)000}+2 \vec{\alpha} _{1(b-3)000} \notag\\ & -6 \vec{\alpha} _{1(b-2)000}+2 b \vec{\alpha} _{1(b-2)000}+2 D \vec{\alpha} _{1(b-2)000}-\vec{\beta} _{12;0(b-3)000}+\vec{\beta} _{13;0(b-3)000}+\vec{\beta} _{23;0(b-3)000})\notag\\ & -s_0^2 (3 \vec{\alpha} _{0(b-3)000}+8 (b-2) \vec{\alpha} _{0(b-2)000}+18 \vec{\alpha} _{0(b-1)000}-26 b \vec{\alpha}
	_{0(b-1)000}+8 b^2 \vec{\alpha} _{0(b-1)000} \notag\\ & -4 D \vec{\alpha} _{0(b-1)000}+4 b D \vec{\alpha} _{0(b-1)000}+4 \vec{\alpha} _{1(b-3)000}-8 \vec{\alpha} _{1(b-2)000}+4 b \vec{\alpha} _{1(b-2)000}-2 D \vec{\alpha} _{1(b-2)000} \notag\\ & +24
	\vec{\alpha} _{1(b-1)000}-32 b \vec{\alpha} _{1(b-1)000}+8 b^2 \vec{\alpha} _{1(b-1)000}-10 D \vec{\alpha} _{1(b-1)000}+10 b D \vec{\alpha} _{1(b-1)000}\notag\\ & -2 \vec{\beta} _{12;0(b-3)000}-2 \vec{\beta} _{12;0(b-3)001} +2 D
	\vec{\beta} _{12;0(b-3)001}+9 \vec{\beta} _{12;0(b-2)000} -4 b \vec{\beta} _{12;0(b-2)000} \notag\\ & -2 D \vec{\beta} _{12;0(b-2)000}+2 \vec{\beta} _{12;1(b-2)000}-2 D \vec{\beta} _{12;1(b-2)000}+2 \vec{\beta} _{13;0(b-3)000}+2 \vec{\beta} _{13;0(b-3)001} \notag\\ & -2 D \vec{\beta} _{13;0(b-3)001}-9 \vec{\beta} _{13;0(b-2)000}+4 b \vec{\beta} _{13;0(b-2)000}+2 D \vec{\beta} _{13;0(b-2)000}-2 \vec{\beta} _{13;1(b-2)000} \notag\\ & +2 D \vec{\beta} _{13;1(b-2)000}+2 \vec{\beta} _{23;0(b-3)000}+2 \vec{\beta} _{23;0(b-3)001}-2 D \vec{\beta} _{23;0(b-3)001}+8 \vec{\beta} _{23;0(b-3)010}\notag\\ & -8 D \vec{\beta} _{23;0(b-3)010}-9 \vec{\beta} _{23;0(b-2)000}+4 b \vec{\beta} _{23;0(b-2)000}+2 D \vec{\beta} _{23;0(b-2)000}+4 \vec{\beta} _{23;0(b-1)000} \notag\\ &-4 b \vec{\beta} _{23;0(b-1)000}-4 D \vec{\beta} _{23;0(b-1)000}+4 b D \vec{\beta} _{23;0(b-1)000}-2 \vec{\beta} _{23;1(b-2)000}+2 D \vec{\beta} _{23;1(b-2)000}))\notag\\ & +s_0 (-M_3^4 (\vec{\alpha} _{0(b-3)000}+(-3+2 b) \vec{\alpha}
	_{0(b-2)000})+M_3^2 s_0 (2 \vec{\alpha} _{0(b-3)000}+(-6+4 b-2 D) \vec{\alpha} _{0(b-2)000} \notag\\ & +6 \vec{\alpha} _{0(b-1)000}-10 b \vec{\alpha} _{0(b-1)000}+4 b^2 \vec{\alpha} _{0(b-1)000}+2 \vec{\alpha} _{1(b-3)000}-10
	\vec{\alpha} _{1(b-2)000}+6 b \vec{\alpha} _{1(b-2)000}\notag\\ & +8 \vec{\alpha} _{1(b-1)000}-12 b \vec{\alpha} _{1(b-1)000}+4 b^2 \vec{\alpha} _{1(b-1)000}-2 D \vec{\alpha} _{1(b-1)000}+2 b D \vec{\alpha} _{1(b-1)000}-\vec{\beta} _{12;0(b-3)000} \notag\\ & +3 \vec{\beta} _{12;0(b-2)000}-2 b \vec{\beta} _{12;0(b-2)000}+\vec{\beta} _{13;0(b-3)000}-3 \vec{\beta} _{13;0(b-2)000}+2 b \vec{\beta} _{13;0(b-2)000}+\vec{\beta} _{23;0(b-3)000}\notag\\ & -3 \vec{\beta} _{23;0(b-2)000}+2 b \vec{\beta} _{23;0(b-2)000})-s_0^2 (\vec{\alpha} _{0(b-3)000}+(-3+2 b-2 D) \vec{\alpha} _{0(b-2)000}-2 \vec{\alpha} _{0(b-1)000} \notag\\ & +2 b \vec{\alpha}
	_{0(b-1)000}+4 D \vec{\alpha} _{0(b-1)000}-4 b D \vec{\alpha} _{0(b-1)000}+2 \vec{\alpha} _{1(b-3)000}-2 \vec{\alpha} _{1(b-2)000}+2 b \vec{\alpha} _{1(b-2)000} \notag\\ & -4 D \vec{\alpha} _{1(b-2)000}+2 D \vec{\alpha} _{1(b-1)000}-2 b D
	\vec{\alpha} _{1(b-1)000} -\vec{\beta} _{12;0(b-3)000}-2 \vec{\beta} _{12;0(b-3)001}+2 D \vec{\beta} _{12;0(b-3)001}\notag\\ & -\vec{\beta} _{12;0(b-2)000}+2 D \vec{\beta} _{12;0(b-2)000}+6 \vec{\beta} _{12;0(b-2)001}-4 b \vec{\beta} _{12;0(b-2)001}-6 D \vec{\beta} _{12;0(b-2)001} \notag\\ & +4 b D \vec{\beta} _{12;0(b-2)001}-2 \vec{\beta} _{12;1(b-1)000}+2 b \vec{\beta} _{12;1(b-1)000}+2 D \vec{\beta} _{12;1(b-1)000}-2 b D \vec{\beta} _{12;1(b-1)000} \notag\\ & +\vec{\beta} _{13;0(b-3)000}+2 \vec{\beta} _{13;0(b-3)001}-2 D \vec{\beta} _{13;0(b-3)001}+\vec{\beta} _{13;0(b-2)000}-2 D \vec{\beta} _{13;0(b-2)000} \notag\\ & -6 \vec{\beta} _{13;0(b-2)001}+4 b \vec{\beta} _{13;0(b-2)001}+6 D \vec{\beta} _{13;0(b-2)001}-4 b D \vec{\beta} _{13;0(b-2)001}+20 \vec{\beta} _{13;1(b-2)000} \notag\\ & -8 b \vec{\beta} _{13;1(b-2)000}-8 D \vec{\beta} _{13;1(b-2)000}+2 \vec{\beta} _{13;1(b-1)000}-2 b \vec{\beta} _{13;1(b-1)000}-2 D \vec{\beta} _{13;1(b-1)000} \notag\\ & +2 b D \vec{\beta} _{13;1(b-1)000}+8 \vec{\beta} _{13;2(b-2)000}-8 D \vec{\beta} _{13;2(b-2)000}+\vec{\beta} _{23;0(b-3)000}+2 \vec{\beta} _{23;0(b-3)001} \notag\\ & -2 D \vec{\beta} _{23;0(b-3)001}+8 \vec{\beta} _{23;0(b-3)010}-8 D \vec{\beta} _{23;0(b-3)010}+\vec{\beta} _{23;0(b-2)000}-2 D \vec{\beta} _{23;0(b-2)000} \notag\\ & -6 \vec{\beta} _{23;0(b-2)001}+4 b \vec{\beta} _{23;0(b-2)001}+6 D \vec{\beta} _{23;0(b-2)001}-4 b D \vec{\beta} _{23;0(b-2)001}-24 \vec{\beta} _{23;0(b-2)010} \notag\\ & +16 b \vec{\beta} _{23;0(b-2)010}+24 D \vec{\beta} _{23;0(b-2)010}-16 b D \vec{\beta} _{23;0(b-2)010}+4 \vec{\beta} _{23;0(b-1)000} -4 b \vec{\beta} _{23;0(b-1)000} \notag\\ & -4 D \vec{\beta} _{23;0(b-1)000}+4 b D \vec{\beta} _{23;0(b-1)000}+8 b \vec{\beta} _{23;0b000}-8 b^2 \vec{\beta} _{23;0b000}-8 b D \vec{\beta} _{23;0b000} \notag\\ & +8 b^2 D \vec{\beta} _{23;0b000}+2 \vec{\beta} _{23;1(b-1)000}-2 b \vec{\beta} _{23;1(b-1)000}-2 D \vec{\beta} _{23;1(b-1)000}+2 b D \vec{\beta} _{23;1(b-1)000}))\}\,.
\end{align}
These expressions for $\vec{\alpha} _{0 b000}$ and $\vec{\alpha} _{1 b000}$ are also invalid when $b=0,1$,  because the denominator becomes zero. This corresponds to the four initial conditions we require: $\vec{\alpha} _{0 0000}$, $\vec{\alpha} _{01000}$, $\vec{\alpha} _{10000}$, $\vec{\alpha} _{11000}$.

	\section{The massless case}\label{sec:massless}
	When one or more masses vanish, the previously derived syzygy equations \eqref{2.15} undergo changes: aside from the seven recurrence relations established earlier, which remain valid, additional independent relations emerge. These new relations help reduce the number of top-level master integrals. Our computational results show that for each vanishing mass, a new relation can be found, reducing the number of top-level master integrals by one. In these cases, we need to modify the previous initial conditions to achieve the simplest reduction. We next present three examples to illustrate this approach.

	\subsection{$M_1=  M_2=M_3= 0$}
	When the masses $M_1=  M_2=M_3= 0$, the master integral basis reduces to a single element $I_1$, which is the simplest case. First, the integrals $I_5, I_6$ and $I_7$ vanish as scaleless integrals, thus we can directly set $\vec{\beta}_{ij}=0$ directly. Second, the integrals $I_2, I_3$ and $I_4$ can be further reduced to expressions in terms of $I_1$.  We select an appropriate solution from the massless syzygy equations to demonstrate this point and derive the corresponding results. The chosen vectors are:
	\begin{align}
		v_1^\mu &= -l_2 \cdot K l_1^\mu -l_1 \cdot K l_2^\mu + l_1 \cdot l_2 K^\mu \,,\notag \\
		v_2^\mu &= -(K^2 +2 l_2 \cdot K) l_2^\mu  +  l^2_2 K^\mu \,.
	\end{align}
	
	Repeating the previous procedure, we substitute this set of vectors into the IBP equations, derive the differential equation for $\vec{\alpha}$, and obtain the recurrence relations among the coefficients $\vec{\alpha}_{abnmk}$. Simultaneously setting $\vec{\beta}_{ij}= 0$ and $n=m=k=0$, the final result is as follows:
    \begin{align}
&s_0 (-2 (b+1)^2 \vec{\alpha} _{a(b+1)000}-a (b+1) \vec{\alpha} _{a(b+1)000}+8 (b+1) \vec{\alpha} _{a(b+1)000}-\vec{\alpha} _{(a-1)b001}-b \vec{\alpha} _{ab000}\notag\\ & +2
\vec{\alpha} _{ab000}-3 (b+1) D \vec{\alpha} _{a(b+1)000}+D \vec{\alpha} _{(a-1)b001}-D \vec{\alpha} _{ab000})=0
    \end{align}
Setting the parameters $a$ and $b$ in this identity to (0,0) (1,0) and (0,1), respectively, and employing the seven recurrence relations obtained previously, we get the following three equations:
\begin{align}
& (D-2) s_0 (\vec{\alpha} _{00000}+3 \vec{\alpha} _{01000})=0\,, \notag \\
& s_0 (2 ((D-1) \vec{\alpha} _{10000}+3 D \vec{\alpha} _{11000}+\vec{\alpha} _{01000}-4 \vec{\alpha} _{11000})+\vec{\alpha} _{00000})=0\,, \notag \\
&s_0 ((3 D-4) \vec{\alpha} _{00000}+(6 D-8) \vec{\alpha} _{10000}+7 D \vec{\alpha} _{01000}+12 D \vec{\alpha} _{11000}-10 \vec{\alpha} _{01000}-16 \vec{\alpha} _{11000})=0\,.
\end{align}
Solving these three equations yields the following results:
\begin{align}
	\vec{\alpha} _{10000}= -\frac{\vec{\alpha} _{00000}}{3}\,, \vec{\alpha} _{01000}= -\frac{\vec{\alpha} _{00000}}{3} \,,\vec{\alpha} _{11000}= \frac{(2 D-3) \vec{\alpha} _{00000}}{6 (3 D-4)}\,.
\end{align}
These are precisely the relations we are looking for:
\begin{align}
	I_2= -\frac{1}{3}I_1\,, I_3= -\frac{1}{3}I_1 \,, I_4= \frac{(2 D-3) }{6 (3 D-4)}I_1\,. 
\end{align}

To summarize, in this massless case, by employing the seven recurrence relations derived previously, setting the masses and $\vec{\beta}_{ij}$  to zero, and taking the initial conditions as: $\vec{\alpha} _{0 0000}=(1,0,0,0,0,0,0)$, $\vec{\alpha} _{10000}=(-\frac{1}{3},0,0,0,0,0,0)$, $\vec{\alpha} _{01000}=(-\frac{1}{3},0,0,0,0,0,0)$, $\vec{\alpha} _{11000}=(\frac{(2 D-3) }{6 (3 D-4)},0,0,0,0,0,0)$, one can compute the  reduction coefficients for the tensor integrals of the sunset topology.

	\subsection{$M_1=  M_2= 0$}
	The case is the same as before. A suitable solution to the syzygy equations can be found to derive new relations. The chosen vectors are:
	\begin{align}
		v_1^\mu &= l_2 \cdot K l_1^\mu +l_1 \cdot K l_2^\mu - l_1 \cdot l_2 K^\mu \,,\notag \\
		v_2^\mu &= -l_2 \cdot K l_1^\mu -l_1 \cdot K l_2^\mu + l_1 \cdot l_2 K^\mu \,.
	\end{align}
	The independent relations we can finally obtain are:
	\begin{align}
		& (D-2) s_0 (\vec{\alpha} _{10000}- \vec{\alpha} _{01000})=0\,, \notag \\
		&s_0 (-D \vec{\alpha} _{0 0000}-2 D \vec{\alpha} _{10000}-3
		D \vec{\alpha} _{01000}-6 D \vec{\alpha} _{11000} +\vec{\alpha} _{0 0000}+2 \vec{\alpha} _{10000}+4 \vec{\alpha} _{01000}+8 \vec{\alpha} _{11000})\notag\\ &+M_3^2 ((D-1) \vec{\alpha} _{0 0000}+(D-2) \vec{\alpha} _{01000})=0\,.
	\end{align}
	If we choose $I_1$ and $I_2$ as the master integrals at this stage, then $I_3$	and $I_4$ can be expressed as
	\begin{align}
		&I_3 = I_2\,, \notag \\
		&I_4 = \frac{D M_3^2-D s_0-M_3^2+s_0}{2 (3 D-4) s_0} I_1  + \frac{D M_3^2-5 D s_0-2 M_3^2+6 s_0}{2 (3 D-4) s_0}I_2 \,.
	\end{align}
	
	Similarly, in this case, by employing the seven recurrence relations derived previously, setting $M_1,  M_2$ and $\vec{\beta}_{ij}$  to zero, and taking the initial conditions as: $\vec{\alpha} _{0 0000}=(1,0,0,0,0,0,0)$, $\vec{\alpha} _{10000}=(0,1,0,0,0,0,0)$, $\vec{\alpha} _{01000}=(0,1,0,0,0,0,0)$ and  $\vec{\alpha} _{11000}=(\frac{D M_3^2-D s_0-M_3^2+s_0}{2 (3 D-4) s_0},\frac{D M_3^2-5 D s_0-2 M_3^2+6 s_0}{2 (3 D-4) s_0},0,0,0,0,0)$, one can compute the  reduction coefficients.

	\subsection{$M_1=  0$}
	The situation changes slightly in this case, but the approach remains unchanged. Master integrals $I_5$ and $I_6$ vanish, while master integral $I_7$ does not. The IBP equation provides only one independent relation. We select $I_1$, $I_2$, $I_3$ and $I_7$ as master integrals and reduce $I_4$ by this pair of vectors:
\begin{align}
	v_1^\mu &= (-l_1 \cdot K l_2^2 +(l_2 \cdot K)^2-  K^2 l_2^2)l_1^\mu +l_1 \cdot K l_2 \cdot K l_2^\mu + (l_1^2 l_2^2- l_1 \cdot l_2 l_2 \cdot K) K^\mu \,,\notag \\
	v_2^\mu &= (-l_2 \cdot K l_2^2 - (l_2 \cdot K)^2)l_1^\mu -l_1 \cdot K l_2 \cdot K l_2^\mu + (l_1 \cdot l_2 l_2^2 +l_1 \cdot K l_2^2 + l_1 \cdot l_2 l_2 \cdot K) K^\mu\,.
\end{align}
The sole novel relation obtained from this IBP equation is:
	\begin{align}
	&\vec{\alpha} _{11000}=\frac{(D-1) \left(M_3^2-s_0\right)+(5-3 D) M_2^2}{2 (3 D-4) s_0}\vec{\alpha} _{00000} +\frac{(3-2 D) M_2^2-(D-1) s_0}{(3 D-4) s_0}\vec{\alpha} _{10000}\notag\\ & +\frac{D M_3^2-D M_2^2-3 D s_0+2 M_2^2-2 M_3^2+4 s_0}{2 (3 D-4) s_0}\vec{\alpha} _{01000}+\frac{1-D}{6 D-8}\vec{\beta} _{23;00000}+\frac{2-D}{6 D-8} \vec{\beta} _{23;01000} \,,
\end{align}
which means:
	\begin{align}
	I_4=&\frac{(D-1) \left(M_3^2-s_0\right)+(5-3 D) M_2^2}{2 (3 D-4) s_0}I_1 +\frac{(3-2 D) M_2^2-(D-1) s_0}{(3 D-4) s_0}I_2\notag\\ & +\frac{D M_3^2-D M_2^2-3 D s_0+2 M_2^2-2 M_3^2+4 s_0}{2 (3 D-4) s_0}I_3+\frac{1-D }{2 (3 D-4) }I_7 \,.
\end{align}

Similarly, by employing the seven recurrence relations derived previously, setting $M_1$, $\vec{\beta}_{12}$ and $\vec{\beta}_{13}$  to zero, and taking the initial conditions as: $\vec{\alpha} _{0 0000}=(1,0,0,0,0,0,0)$, $\vec{\alpha} _{10000}=(0,1,0,0,0,0,0)$, $\vec{\alpha} _{01000}=(0,0,1,0,0,0,0)$ and  $\vec{\alpha} _{11000}=(\frac{(D-1) \left(M_3^2-s_0\right)+(5-3 D) M_2^2}{2 (3 D-4) s_0}$,\\$\frac{(3-2 D) M_2^2-(D-1) s_0}{(3 D-4) s_0}, \frac{D M_3^2-d M_2^2-3 D s_0+2 M_2^2-2 M_3^2+4 s_0}{2 (3 D-4) s_0},0,0,0,\frac{1-D }{2 (3 D-4) })$, one can compute the  reduction coefficients.

\section{\label{sec:conc}Conclusion}

In this paper, we have studied the tensor reduction of sunset diagrams using the generating function with the auxiliary vector $R_i$. By combining the PV-reduction method and the syzygy method, we construct a complete system of differential equations for the generating function. Through the series expansion of the reduction coefficients, differential equations give a system of relations for the reduction coefficients. Solving them we derive recurrence relations for the reduction coefficients, expressing higher-order terms in terms of lower-order ones.

The sunset is just a demonstration of the applicability of generating functions to the reduction of higher loop integrals. A natural thing to do is to apply this method to other topologies as well as formulate  a systematic algorithm  based on the experience of this work.

 \paragraph{Note added:} After the release of our paper, the paper \cite{Feng:2025leo} appeared. In this paper, different generating functions have been used, where the expansion coefficients are familiar Feynman integrals in the family. The technique used in \cite{Feng:2025leo}, i.e., solving the highest degree differential operator, could simplify the solving procedure presented in our paper. We will pursue it in another future project.

\section*{Acknowledgement}
 This work is supported by the National Natural Science Foundation of China through Grants No. 12535003, No.11935013, No.11947301, and No.12047502.

\appendix

\section{ Tadpole result}\label{sec:Tadpole result}
 For the one-loop rank $r$ tensor tadpole, the reduction is \cite{Hu:2024kch,Feng:2022hyg} given by
\begin{align}
	I_1^{(r)}\equiv \int \frac{d^D l}{i \pi^{D/2}} \frac{(R\cdot l)^r}{(l-K)^2 - M^2_0} \equiv \alpha^{(tad)}[K,M_0;R;r]\int \frac{d^D l}{i \pi^{D/2}} \frac{1}{(l-K)^2 - M^2_0}\,,
\end{align}
where the reduction coefficient $\alpha^{(tad)}[K,M_0;r]$  is well known. For the case $K=0$, we have
\begin{align}
\alpha^{(tad)}[0,M_0;R;r] =  \frac{1+(-1)^r}{2} \frac{(r-1)!!M_0^r}{\prod_{i=1}^{\frac{r}{2}}(D + 2(i-1))}(R^2)^{\frac{r}{2}}\,.
\end{align}
Introducing  the Pochhammer symbol, it can be written as 
\begin{align}
\alpha^{(tad)}[0,M_0;R;r] =  \frac{1+(-1)^r}{2} \frac{(r-1)!!M_0^r}{2^{\frac{r}{2}} (\frac{D}{2})_{\frac{r}{2}}}(R^2)^{\frac{r}{2}}\,.
\end{align}
 For the case $K \neq 0$, we write
\begin{align}
&\int \frac{d^D l}{i \pi^{D/2}} \frac{(R \cdot l)^r}{(l-K)^2 - M^2_0} = \int \frac{d^D l}{i \pi^{D/2}} \frac{(R \cdot (l + K))^r}{l^2 - M^2_0} \notag \\ =&\sum^{r}_{t=0} \frac{r!}{t! (r - t)!} (R\cdot K)^{r-t}\int \frac{d^D l}{i \pi^{D/2}} \frac{(R\cdot l)^t}{l^2 - M^2_0}\,,
\end{align}
 thus we have
\begin{align}
\alpha^{(tad)}[K,M_0;R;r] = \sum^{r}_{t=0} \frac{1+(-1)^t}{2} \frac{r!}{t! (r - t)!} \frac{(t-1)!!M^t_0}{2^{\frac{t}{2}} (\frac{D}{2})_{\frac{t}{2}}}(R \cdot K)^{r-t}(R^2)^{\frac{t}{2}}\,.
\end{align}                            
This result can be written as
\begin{align}
	\alpha^{(tad)}[K,M_0;R;r] &= \sum_{r_1 + 2r_2 = r}\frac{(r_1 + 2r_2)!(2r_2 - 1)!!M_0^{2r_2}}{(2r_2)!r_1 ! 2^{r_2}(\frac{D}{2})_{r_2}}(R\cdot K)^{r_1}(R^2)^{r_2}\notag\\&=\sum_{r_1 + 2r_2 = r}\frac{(r_1 + 2r_2)! M_0^{2r_2}}{r_2!r_1 ! 2^{2r_2}(\frac{D}{2})_{r_2}}(R\cdot K)^{r_1}(R^2)^{r_2}\,,
\end{align}
 where the sum is over all non-negative integers of $r_1$,$r_2$ with constraint $r_1+2r_2=r$.
 
The above analysis is a little bit complicated. Using the generation function, we have
\begin{align}
\int dl\frac{e^{l \cdot R}}{l^2 - M^2} = \beta(R,M)\int d l\frac{1}{l^2 - M^2},~~~~~\beta(R,M) = \sum^{\infty}_{n=0} \frac{(\frac{1}{4}M^2 R^2)^n}{n!(\frac{D}{2})_n}\,.
\end{align}
 Using it, we can see that
\begin{align}
	&\int \frac{d^D l_1}{i \pi^{D/2}}\frac{d^D l_2}{i \pi^{D/2}}\frac{e ^ {l_1 \cdot R_1} e ^ {l_2 \cdot R_2}}{D_2 D_3}=\int \frac{d^D l_1}{i \pi^{D/2}}\frac{d^D l_2}{i \pi^{D/2}}\frac{e ^ {l_1 \cdot R_1} e ^ {l_2 \cdot R_2}}{(l^2_2 - M^2_2) ((l_1 + l_2 + K)^2 - M^2_3)}\notag \\ =&\int \frac{d^D l_1}{i \pi^{D/2}}\frac{d^D l_2}{i \pi^{D/2}}\frac{e ^ {(l_1 - l_2 - K)\cdot R_1} e ^ {l_2 \cdot R_2}}{(l^2_2 - M^2_2) (l_1^2 - M^2_3)} =  K^2 \beta_{23} \frac{1}{K^2}	\int \frac{d^D l_1}{i \pi^{D/2}}\frac{d^D l_2}{i \pi^{D/2}}\frac{1}{D_2 D_3}\,,
\end{align}
 with
\begin{align}
	\beta_{23} = e^{-K\cdot R_1} \beta(R_2 - R_1,M_2)\beta(R_1,M_3)\,.
\end{align}
 Similarly we have
\begin{align}
\int \frac{d^D l_1}{i \pi^{D/2}}\frac{d^D l_2}{i \pi^{D/2}}\frac{e ^ {l_1 \cdot R_1} e ^ {l_2 \cdot R_2}}{D_1 D_2} = K^2 \beta_{12} \frac{1}{K^2}	\int \frac{d^D l_1}{i \pi^{D/2}}\frac{d^D l_2}{i \pi^{D/2}}\frac{1}{D_1 D_2}\,,
\end{align}
 with
\begin{align}
	\beta_{12} =  \beta(R_1 ,M_1)\beta(R_2,M_2)\,,
\end{align}
and
\begin{align}
	\int \frac{d^D l_1}{i \pi^{D/2}}\frac{d^D l_2}{i \pi^{D/2}}\frac{e ^ {l_1 \cdot R_1} e ^ {l_2 \cdot R_2}}{D_1 D_3} = K^2 \beta_{13} \frac{1}{K^2}	\int \frac{d^D l_1}{i \pi^{D/2}}\frac{d^D l_2}{i \pi^{D/2}}\frac{1}{D_1 D_3}\,,
\end{align}
with
\begin{align}
	\beta_{13} = e^{-K\cdot R_2} \beta(R_1 - R_2,M_1)\beta(R_2,M_3)\,.
\end{align}
 The choice makes the $\beta_{ij}$ mass dimensionless.

\section{The rank-four integral}\label{sec:example}
Here we present an  example for a rank-four tensor reduction to illustrate the application of our result. As the final result is rather lengthy(spanning approximately four pages), we set $M_1= 13, M_2= 3, M_3= 7, s_0= 11, D= 5, H^{\mu \nu }=s_0 g^{\mu \nu}-K^{\nu} K^{\mu} $  to simplify the expression. Our result is in agreement with that obtained from FIRE6 \cite{Smirnov:2019qkx,Smirnov:2023yhb}.
\begin{align}
	\int &\frac{d^D l_1}{i \pi^{D/2}}\frac{d^D l_2}{i \pi^{D/2}}\frac{l^{\mu}_1 l^{\nu}_1 l^{\alpha}_2 l^{\beta}_2}{D_1 D_2 D_3} = \frac{\partial}{\partial R^{\mu}_1}\frac{\partial}{\partial R^{\nu}_1}\frac{\partial}{\partial R^{\alpha}_2}\frac{\partial}{\partial R^{\beta}_2}\int \frac{d^D l_1}{i \pi^{D/2}}\frac{d^D l_2}{i \pi^{D/2}}\frac{e ^ {l_1 \cdot R_1} e ^ {l_2 \cdot R_2}}{D_1 D_2 D_3}\Big|_{R_1=0,R_2=0} \notag \\ =& \vec{I} \cdot \{\sum_{abnmk} \vec{\alpha}_{abnmk} \frac{\partial}{\partial R^{\mu}_1}\frac{\partial}{\partial R^{\nu}_1}\frac{\partial}{\partial R^{\alpha}_2}\frac{\partial}{\partial R^{\beta}_2} (y^a_1 y^b_2 x^n_{11} x^m_{22} x^k_{12})\}\Big|_{R_1=0,R_2=0}
	\notag \\ =& \vec{I}\cdot\{4K^{\mu}K^{\nu}K^{\alpha}K^{\beta}\vec{\alpha}_{22000} + 4 H^{\mu\nu}K^{\alpha}K^{\beta}\vec{\alpha}_{02100} +4 H^{\alpha\beta}K^{\mu}K^{\nu}\vec{\alpha}_{20010} \notag \\ &+ 4 H^{\alpha\beta}H^{\mu\nu}\vec{\alpha}_{00110} + 2(H^{\alpha\mu}H^{\beta\nu} + H^{\alpha\nu}H^{\mu\beta})\vec{\alpha}_{00002} \notag \\ &+ (H^{\alpha\mu}K^{\beta}K^{\nu}+H^{\beta\mu}K^{\alpha}K^{\nu}+H^{\alpha\nu}K^{\mu}K^{\beta}+H^{\beta\nu}K^{\mu}K^{\alpha})\vec{\alpha}_{11001}\} 
	     \notag \\ =& \frac{1}{622908}({80280993 K^{\alpha} K^{\beta} H^{\mu \nu }+94695393 K^{\mu} K^{\nu} H^{\alpha \beta }+89245245 (K^{\alpha} K^{\mu} H^{\beta \nu } }\notag \\ & + {K^{\beta} K^{\mu} H^{\alpha \nu }+K^{\nu} (K^{\alpha} H^{\mu \beta }+K^{\beta} H^{\mu \alpha }))-382024812 K^{\alpha} K^{\beta} K^{\mu} K^{\nu}}\notag \\ & - {12791297 H^{\alpha \beta } H^{\mu \nu }-13173734 (H^{\nu \alpha } H^{\mu \beta }+H^{\mu \alpha } H^{\nu \beta })})\int \frac{d^D l_1}{i \pi^{D/2}}\frac{d^D l_2}{i \pi^{D/2}}  \frac{1}{D_1 D_2 D_3}  
	\notag \\ & +\frac{1}{13703976}({42648060 K^{\alpha} K^{\beta} H^{\mu \nu }+49471734 K^{\mu} K^{\nu} H^{\alpha \beta }+47247600 (K^{\alpha} K^{\mu} H^{\beta \nu }}\notag \\ &  +{K^{\beta}K^{\mu} H^{\alpha \nu }+K^{\nu} (K^{\alpha} H^{\mu \beta }+K^{\beta} H^{\mu \alpha }))-205392720 K^{\alpha} K^{\beta} K^{\mu} K^{\nu}}\notag \\ &  -{6654029
	H^{\alpha \beta } H^{\mu \nu }-7077749 (H^{\nu \alpha } H^{\mu \beta }+H^{\mu \alpha } H^{\nu \beta })}) \int \frac{d^D l_1}{i \pi^{D/2}}\frac{d^D l_2}{i \pi^{D/2}}  \frac{l_1 \cdot K}{D_1 D_2 D_3}
		\notag \\ & +\frac{1}{4567992}({57129618 K^{\alpha} K^{\beta} H^{\mu \nu }+67484820 K^{\mu} K^{\nu} H^{\alpha \beta }+63231120 (K^{\alpha} K^{\mu} H^{\beta \nu }}\notag \\ &  +{K^{\beta}K^{\mu} H^{\alpha \nu }+K^{\nu} (K^{\alpha} H^{\mu \beta }+K^{\beta} H^{\mu \alpha }))-275499120 K^{\alpha} K^{\beta} K^{\mu} K^{\nu}}\notag \\ &  -{8957743
	H^{\alpha \beta } H^{\mu \nu }-9954343 (H^{\nu \alpha } H^{\mu \beta }+H^{\mu \alpha } H^{\nu \beta })})\int \frac{d^D l_1}{i \pi^{D/2}}\frac{d^D l_2}{i \pi^{D/2}}  \frac{l_2 \cdot K}{D_1 D_2 D_3}
		\notag \\ & +\frac{1}{13703976}({4098198 K^{\alpha} K^{\beta} H^{\mu \nu }+4921878 K^{\mu} K^{\nu} H^{\alpha \beta }+4552410 (K^{\alpha} K^{\mu} H^{\beta \nu }}\notag \\ & +{K^{\beta}
	K^{\mu} H^{\alpha \nu }+K^{\nu} (K^{\alpha} H^{\mu \beta }+K^{\beta} H^{\mu \alpha }))-19872840 K^{\alpha} K^{\beta} K^{\mu} K^{\nu}}\notag \\ & -{659603
	H^{\alpha \beta } H^{\mu \nu }-706727 (H^{\nu \alpha } H^{\mu \beta }+H^{\mu \alpha } H^{\nu \beta })})\int \frac{d^D l_1}{i \pi^{D/2}}\frac{d^D l_2}{i \pi^{D/2}} \frac{ ({l_1 \cdot K})({l_2 \cdot K})}{D_1 D_2 D_3} 
	     \notag \\ &+ \frac{1}{12458160}({-11273808 K^{\alpha} K^{\beta} H^{\mu \nu }-13333008 K^{\mu} K^{\nu} H^{\alpha \beta }-12597438 (K^{\alpha} K^{\mu} H^{\beta \nu }}\notag \\ & +{K^{\beta}
	K^{\mu} H^{\alpha \nu }+K^{\nu} (K^{\alpha} H^{\mu \beta }+K^{\beta} H^{\mu \alpha }))+53795352 K^{\alpha} K^{\beta} K^{\mu} K^{\nu}}\notag \\ & +{1862963
	H^{\alpha \beta } H^{\mu \nu }+1853063 (H^{\nu \alpha } H^{\mu \beta }+H^{\mu \alpha } H^{\nu \beta })})\int \frac{d^D l_1}{i \pi^{D/2}}\frac{d^D l_2}{i \pi^{D/2}}  \frac{1}{D_1 D_2 } 
	    \notag \\ &+ \frac{1}{12458160}({4630080 K^{\alpha} K^{\beta} H^{\mu \nu }+5384262 K^{\mu} K^{\nu} H^{\alpha \beta }+4715616 (K^{\alpha} K^{\mu} H^{\beta \nu }}\notag \\ & +{K^{\beta}
	K^{\mu} H^{\alpha \nu }+K^{\nu} (K^{\alpha} H^{\mu \beta }+K^{\beta} H^{\mu \alpha }))-18520320 K^{\alpha} K^{\beta} K^{\mu} K^{\nu}}\notag \\ & -{64127
	H^{\alpha \beta } H^{\mu \nu }-388847 (H^{\nu \alpha } H^{\mu \beta }+H^{\mu \alpha } H^{\nu \beta })})\int \frac{d^D l_1}{i \pi^{D/2}} \frac{d^D l_2}{i \pi^{D/2}}  \frac{1}{D_1  D_3}
	   \notag \\ &+ \frac{1}{12458160}({11524422 K^{\alpha} K^{\beta} H^{\mu \nu }+13653120 K^{\mu} K^{\nu} H^{\alpha \beta }+12851616 (K^{\alpha} K^{\mu} H^{\beta \nu }}\notag \\ & +{K^{\beta}
	K^{\mu} H^{\alpha \nu }+K^{\nu} (K^{\alpha} H^{\mu \beta }+K^{\beta} H^{\mu \alpha }))-54612480 K^{\alpha} K^{\beta} K^{\mu} K^{\nu}}\notag \\ & -{1879487
	H^{\alpha \beta } H^{\mu \nu }-1887407 (H^{\nu \alpha } H^{\mu \beta }+H^{\mu \alpha } H^{\nu \beta })})\int \frac{d^D l_1}{i \pi^{D/2}}\frac{d^D l_2}{i \pi^{D/2}}  \frac{1}{ D_2 D_3}\,,
\end{align}

For the massless case $(M_1=  M_3=M_3= 0)$, the result is relatively simple, and we present its complete expression as follows:
\begin{align}
	\int &\frac{d^D l_1}{i \pi^{D/2}}\frac{d^D l_2}{i \pi^{D/2}}\frac{l^{\mu}_1 l^{\nu}_1 l^{\alpha}_2 l^{\beta}_2}{D_1 D_2 D_3}= \frac{1}{36 (D-1) (3 D-4) (3 D-2)}\{-4 D^2 K^{\alpha} K^{\beta} H^{\mu \nu }+2 D^2 K^{\alpha} K^{\mu} H^{\beta \nu }\notag \\ & +2 D^2 K^{\beta} K^{\mu} H^{\alpha \nu }+2 D^2 K^{\beta}
		K^{\nu} H^{\mu \alpha }+H^{\alpha \beta } ((4 D-6) H^{\mu \nu }-2 \left(2 D^2-4 D+3\right) K^{\mu} K^{\nu})\notag \\ & +8 D K^{\alpha} K^{\beta
	} H^{\mu \nu }-D K^{\alpha} K^{\mu} H^{\beta \nu }-D K^{\beta} K^{\mu} H^{\alpha \nu }-D K^{\beta} K^{\nu} H^{\mu \alpha }+D H^{\mu \beta }
		((2 D-1) K^{\alpha} K^{\nu} \notag \\ & +H^{\nu \alpha })-6 K^{\alpha} K^{\beta} H^{\mu \nu }+4 D^3 K^{\alpha} K^{\beta} K^{\mu} K^{\nu}-12 D^2 K^{\alpha}
		K^{\beta} K^{\mu} K^{\nu}+14 D K^{\alpha} K^{\beta} K^{\mu} K^{\nu} \notag \\ & -6 K^{\alpha} K^{\beta} K^{\mu} K^{\nu}+D H^{\mu \alpha } H^{\nu \beta
	}\}\int \frac{d^D l_1}{i \pi^{D/2}}\frac{d^D l_2}{i \pi^{D/2}}  \frac{1}{D_1 D_2 D_3}
\end{align}

\bibliographystyle{JHEP}
\bibliography{reference}

\end{document}